\documentclass[11pt]{article}

\usepackage{authblk}
\usepackage[margin=0.75in]{geometry}

\usepackage{tikz}
\usepackage{amsmath}
\usepackage{filecontents}
\usepackage{subcaption}
\usepackage{fontawesome5}
\usepackage{threeparttable}
\usepackage{amsmath}
\usepackage{tikz,lipsum}
\usepackage{footmisc}
\usepackage{diagbox}
\usepackage[most]{tcolorbox}
\usepackage{enumitem}
\usepackage{placeins}
\usepackage{graphicx}
\usepackage[english]{babel}
\usepackage[colorinlistoftodos]{todonotes}
\usepackage{capt-of}
\usepackage{regexpatch}
\usepackage[font=small,labelfont=bf]{caption}
\DeclareCaptionFont{13pt}{\fontsize{10pt}{10pt}\selectfont}
\usepackage{float}
\usepackage{subcaption}
\usepackage{xspace}
\usepackage{multirow,tabularx}
\usepackage{booktabs}
\usepackage{array} 
\usepackage{xcolor} 
\usepackage{colortbl}
\usepackage{tabularx}
\usepackage[font=small,labelfont=bf]{caption}
\definecolor{aliceblue}{rgb}{0.94, 0.97, 1.0}
\usepackage{stfloats}

\newcolumntype{W}{>{\columncolor{white}}l}
\usepackage{array,ragged2e} 
\newcolumntype{L}[1]{>{\RaggedRight\arraybackslash}p{#1}} 

\usepackage{tcolorbox}
\usepackage{fontspec}
\usepackage{microtype}

\definecolor{headercolor}{RGB}{120,0,0}
\definecolor{inputcolor}{RGB}{230,240,255}
\definecolor{outputcolor}{RGB}{255,230,230}
\definecolor{bracketcolor}{RGB}{0,80,150}

\newtcolorbox{samplebox}{
  enhanced,
  colback=white,
  colframe=headercolor,
  arc=1mm,
  boxrule=0.3pt,
  title={\textcolor{white}{\large\bfseries Input Sample}},
  coltitle=white,
  fonttitle=\bfseries,
  boxsep=0pt,
  toprule=3pt,
  left=2mm,
  right=2mm,
  top=2mm,
  bottom=2mm,
  middle=2mm
}

\newtcolorbox{samplebox2}{
  enhanced,
  colback=white,
  colframe=headercolor,
  arc=1mm,
  boxrule=0.3pt,
  boxsep=0pt,
  left=2mm,
  right=2mm,
  top=2mm,
  bottom=2mm,
  middle=2mm
}

\newtcolorbox{inputbox}{
  enhanced,
  colback=inputcolor,
  colframe=headercolor!40,
  boxrule=0.5pt,
  left=1.5mm,
  right=1.5mm,
  top=1.5mm,
  bottom=1.5mm,
  fontupper=\small,
  boxsep=0pt,
  before skip=0.1mm,
  after skip=2mm,
}

\newtcolorbox{outputbox}{
  enhanced,
  colback=outputcolor,
  colframe=headercolor!40,
  boxrule=0.5pt,
  left=1.5mm,
  right=1.5mm,
  top=1.5mm,
  bottom=1.5mm,
  fontupper=\small,
  boxsep=0pt,
  before skip=2mm,
  after skip=2mm,
}

\newcommand{\revision}[1]{{\leavevmode\color{black}#1}}

\newcommand{\dpguard}[1]{{\leavevmode\textit{#1}}}

\usepackage{hyperref}
\usepackage{cleveref}

\AtBeginDocument{%
  \providecommand\BibTeX{{%
    \normalfont B\kern-0.5em{\scshape i\kern-0.25em b}\kern-0.8em\TeX}}}

\newcommand{\name}{\mbox{\textit{AutoBot}}\xspace}
\newcommand{\textmap}{\mbox{\textit{ElementMap}}\xspace}
\newcommand{\distill}{\mbox{$\mathcal{D_\text{distill}}$}\xspace}

\newcolumntype{Y}{>{\centering\arraybackslash}X} 

\begin{document}

\title{Automatically Detecting Online Deceptive Patterns}

\renewcommand*{\Authsep}{, }
\renewcommand*{\Authand}{, }
\renewcommand*{\Authands}{, }
\renewcommand*{\Affilfont}{\normalsize\normalfont}
\renewcommand*{\Authfont}{\bfseries}    %
\setlength{\affilsep}{2em}   %

\newsavebox\affbox
\newcommand*\samethanks[1][\value{footnote}]{\footnotemark[#1]}
\author{Asmit Nayak}
\author{Shirley Zhang\thanks{Equal Contribution}}
\author{Yash Wani\samethanks}
\author{Rishabh Khandelwal}
\author{Kassem Fawaz}
\affil[]{%
  \savebox\affbox{\Affilfont Department of Chemical Engineering, University of AAAAA BBBBBB, CCCCC road,}%
  \parbox[t]{\wd\affbox}{\protect\centering} University of Wisconsin -- Madison} 
\date{} 

\maketitle

\begin{abstract}

Deceptive patterns in digital interfaces manipulate users into making unintended decisions, exploiting cognitive biases and psychological vulnerabilities. These patterns have become ubiquitous on various digital platforms. While efforts to mitigate deceptive patterns have emerged from legal and technical perspectives, a significant gap remains in creating usable and scalable solutions. We introduce our \name framework to address this gap and help web stakeholders navigate and mitigate online deceptive patterns. \name accurately identifies and localizes deceptive patterns from a screenshot of a website without relying on the underlying HTML code.
\name employs a two-stage pipeline that leverages the capabilities of specialized vision models to analyze website screenshots, identify interactive elements, and extract textual features. Next, using a large language model, \name understands the context surrounding these elements to determine the presence of deceptive patterns.  We also use \name, to create a synthetic dataset to distill knowledge from `\textit{teacher}' LLMs to smaller language models. Through extensive evaluation, we demonstrate \textit{AutoBot}’s effectiveness in detecting deceptive patterns on the web, achieving an F1-score of 0.93 when detecting deceptive patterns, underscoring its potential as an essential tool for mitigating online deceptive patterns.

We implement \name, across three downstream applications targeting different web stakeholders: (1) a local browser extension providing users with real-time feedback, (2) a Lighthouse audit to inform developers of potential deceptive patterns on their sites, and (3) as a measurement tool designed for researchers and regulators.

\end{abstract}



\section{Introduction}
\label{sec:introduction}
\begin{figure}[t]
    \centering
  \includegraphics[width=0.6\columnwidth]{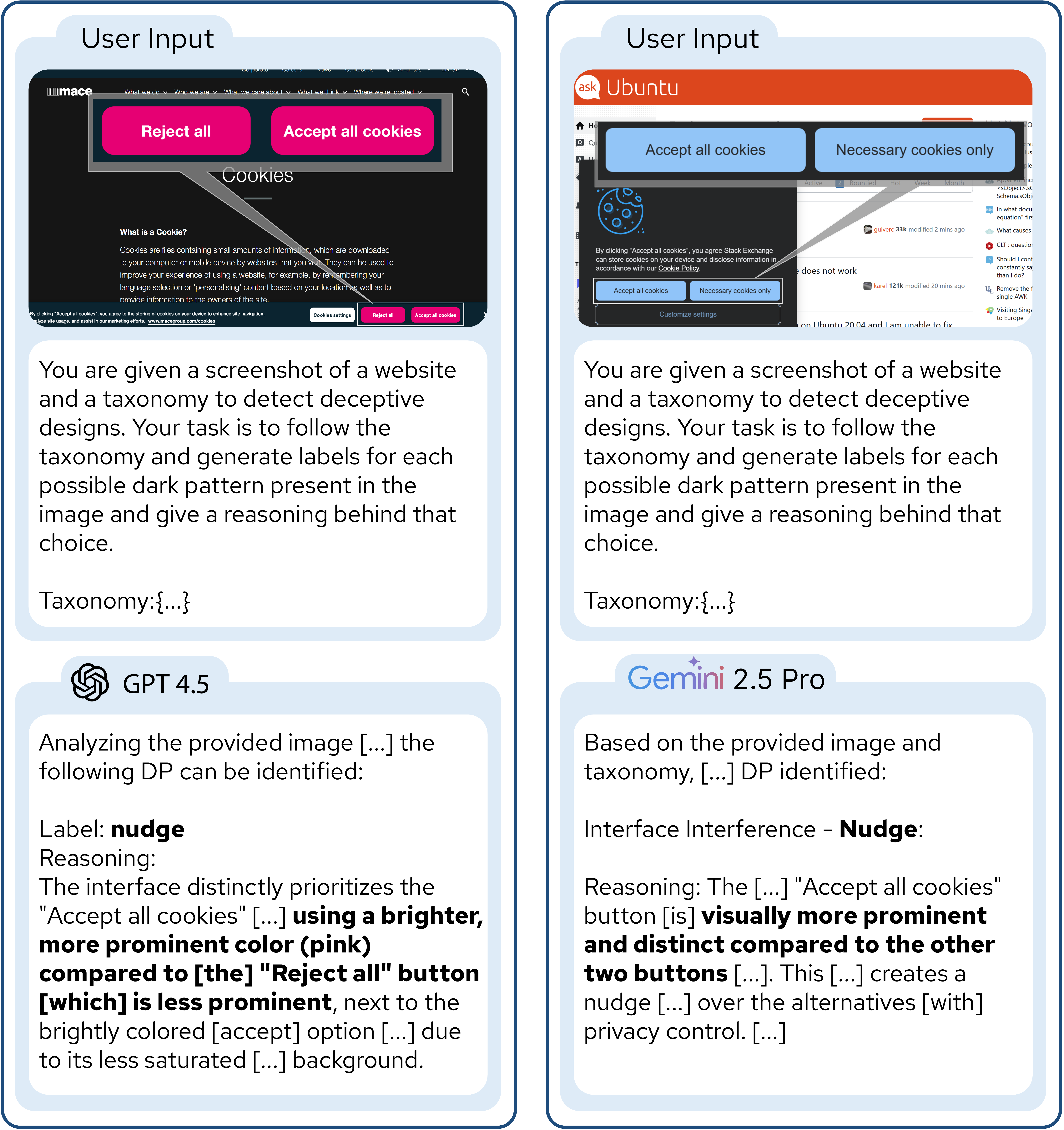}
    \caption {OpenAI's \texttt{GPT4.5} incorrectly identifies the color of the ``Reject All'' button as being less prominent than the other, leading to an incorrect classification of ``nudge''. Similarly, \texttt{Gemini 2.5 Pro} incorrectly notes that the two buttons in the cookie notices are visually distinct from each other, resulting in a misclassification of ``nudge''.}
    \label{fig:gemini_chat}
    
\end{figure}

Deceptive patterns, also known as `\textit{Dark Patterns},' are design choices that manipulate users into making unintended decisions as they interact with applications. These patterns exploit cognitive biases and psychological vulnerabilities to influence user behavior, often in ways that benefit the service provider at the user's expense~\cite{dark_pattern_site, brignull-2023}. 
The growing use of deceptive patterns negatively impacts the quality of user experiences across various online activities, such as online purchases, engaging with social media, playing video games, or simply browsing the web~\cite{mathur2019dark}. The widespread nature of these patterns is well documented, with notable examples from resources like Harry Brignull's \texttt{deceptive.design}\footnote{\url{https://www.deceptive.design/}} and Caroline Sliders' analysis at \texttt{The Pudding}\footnote{\url{https://pudding.cool/2023/05/dark-patterns/}}.

Despite the recent push toward better web experience, fueled by user awareness, media revelations, and privacy regulations~\cite{CPRA, GDPR2016}, deceptive patterns continue to pose substantial challenges~\cite{leiser2023dark}. As a result, users are at risk of harm, such as financial loss~\cite{SchneiderWallace2023}, privacy violations~\cite{Jika2023}, and the exploitation of vulnerable populations, including children~\cite{luke2019}. In response, researchers have explored different approaches to detect and classify deceptive patterns on the web. Earlier efforts included manual analysis to analyze the distribution of deceptive patterns on the Internet~\cite{mathur2019dark,gray2018dark,dark_pattern_site}. Such approaches are infeasible at scale due to the sheer volume, dynamic nature, and diversity of website interfaces. More recent efforts include heuristic- and ML-based methods~\cite{dark_pattern_site,gray2018dark,mathur2019dark, Insite2023,soe2020circumvention,adorna2024developing,chen2023unveiling,raju2022smart,mansur2023aidui}. However, these approaches often exhibit limited accuracy when identifying deceptive patterns in the wild, as we show later.

Recognizing the limitations of current approaches, there is a pressing need for automated tools to assist web stakeholders in navigating and mitigating online deceptive patterns. Such an automated tool must perform two primary tasks: 1) accurately identify deceptive patterns and 2) precisely \textit{localize} their position within the website. The ability to both identify and localize these patterns offers several benefits to web stakeholders. First, web users can be alerted to deceptive patterns on websites they visit, enabling informed decision-making. Second, regulators can leverage such tools to identify deceptive patterns at scale, facilitating enforcement and policy development. Third, developers can gain insights into potentially problematic elements within their websites, promoting more ethical design practices~\cite{stover_how_2023}.

We propose an automated deceptive pattern detection framework, \name, to address these limitations. \name accurately identifies and localizes the deceptive patterns from a screenshot of a webpage. It does not rely on the underlying \texttt{HTML} code of the webpage, which tends to be less stable than screenshots. \texttt{HTML} implementation and code can vary significantly across webpages and even different accesses, while screenshots and text remain more consistent~\cite{khandelwal_prisec_2021}. \name adopts a modular design, breaking down the task into two distinct modules and leveraging existing state-of-the-art models for each.  Specifically, \name utilizes a  \textit{Vision Module}, which analyzes screenshots to accurately localize UI elements, extracting essential features into a structured, text-only format (\textmap). It feeds this \textmap to a \textit{Language Module} that employs a Large Language Model (LLM) to analyze the \textmap and assign a deceptive pattern based on a defined taxonomy (\Cref{subsec:taxonomy}).

\name's design avoids the pitfalls of directly prompting Vision Large Language Model (VLLM) for end-to-end analysis. These models, as we show in~\Cref{fig:gemini_chat}, are known to hallucinate, leading to false positives undermining their reliability~\cite{50Shades}. Furthermore, VLLMs currently lack the capability for accurate localization of UI elements, as shown in recent works~\cite{chen2023shikra,you2023ferret} and in~\Cref{subsec:eval:vision}. While fine-tuning VLLMs could potentially improve performance, the substantial demand for large annotated datasets and significant computing resources renders this approach impractical~\cite{AnthropicHaikuFineTune2024}.

\name leverages the capabilities of specialized vision models to help with the localization task and utilizes LLMs to perform accurate deceptive pattern identification. While LLMs have shown strong performance in detecting deceptive patterns (see~\Cref{subsec:eval:lang}), large-scale use of these LLMs might be prohibitive due to cost, latency, and privacy concerns. In this work, we show how we can create a synthetic dataset using an LLM as the `\textit{teacher}' and distill its knowledge into smaller language models (SLMs), like \texttt{Qwen2.5-1.5B}, and very small language models (vSLMs), like \texttt{Flan-T5}. We show a detailed evaluation of these models in~\Cref{subsec:eval:lang}. 

We demonstrate the practical applications of \name in three instantiations, targeting different web stakeholders. First, we design, and implement a browser extension (\Cref{subsec:browser_extension}) using \name to automatically detect and highlight deceptive patterns on websites, providing real-time user assistance on personal computers. Second, we create a custom Lighthouse audit (\Cref{subsec:lighthouse}) that leverages \name to inform developers of potential deceptive patterns on their sites, integrating directly into developer workflows and providing a quantifiable score. Third, we demonstrate how researchers and regulators can use \name to perform a large-scale measurement and analysis of the online deceptive patterns landscape (\Cref{subsec:measurements}). Our measurement on over 11,000 websites across popular (Tranco) and e-commerce (Shopify) domains highlighted the prevalence of deceptive patterns, with many websites exhibiting several patterns on a single webpage.

\section{Background and Related Works}
\label{sec:background}
Web deceptive patterns refer to website interface design choices that manipulate or deceive users into making decisions they might not otherwise make~\cite{dark_pattern_site}. Examples of such patterns include hidden costs, forced continuity, misdirection in e-commerce websites, and privacy-invasive default settings in social media platforms. Here, we present a filtered taxonomy to categorize web deceptive patterns based on existing work. We also survey existing works on detecting deceptive patterns on websites.

\subsection{Taxonomy for Deceptive Patterns}
\label{subsec:taxonomy}

\begin{figure}[h]
    \centering
  \includegraphics[width=\columnwidth]{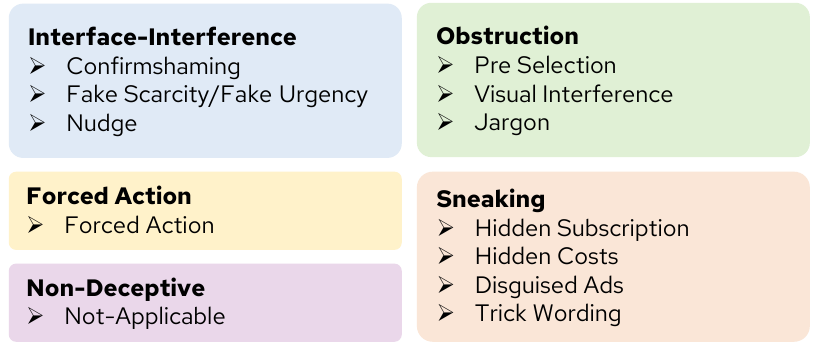}
    \caption{Taxonomy of Deceptive Patterns. \name classifies text elements into five \revision{high-level} deceptive pattern \revision{categories}: Interface Interference, Obstruction, Forced Action, Sneaking, and Non-Deceptive.}
    \label{fig:taxonomy}
\end{figure}

Brignull et al. presented the first taxonomy of deceptive patterns in 2010~\cite{dark_pattern_site}.  Conti et al. expanded this taxonomy to include `malicious interface designs'~\cite{conti2010malicious}.
Bösch et al.~\cite{bosch2016tales} introduced a similar taxonomy called `privacy dark patterns', which included more privacy-centric categories such as `Forced Registration' and `Hidden Legalese Stipulations.'

More recently, Gray et al.~\cite{gray2018dark} created a unified corpus to detect deceptive designs in user interfaces, building on previous taxonomies and categories. Since then, various works have further adapted the taxonomy for specific domains. For instance, Lewis et al.~\cite{lewis_gameful_2014} codified deceptive patterns for mobile apps and games, while Mathur et al.~\cite{mathur2019dark} adapted the taxonomy to focus on deceptive patterns present in shopping websites. Work by Chen et al.~\cite{chen2023unveiling} and Mansur et al.~\cite{mansur2023aidui} extended the taxonomy to detect deceptive patterns in mobile and web apps. Although prior works developed various taxonomies to classify deceptive patterns, these efforts are inconsistent and often domain-specific. To address these issues, Gray et al.~\cite{gray2023towards} introduced a unified ontology of deceptive patterns integrating past literature across regulatory reports and academic works.

\subsubsection{Filtered Taxonomy} 
Since our work, like others in literature, focuses on detecting patterns at the page level, some patterns fall outside its scope. For instance, the deceptive pattern `\textit{Sneak into Basket}' -- where websites quietly add unwanted items like magazine subscriptions to a user's cart -- creates an unwanted transaction that would require the detection system to understand user action across multiple pages over a period of time.

\revision{We classify deceptive patterns into two types -- (1) static patterns with no temporal dimension and are visible upon page render, and (2) dynamic patterns that rely on user interactions or time-based triggers related (e.g., ‘Nagging,’ ‘Hard-to-Cancel,’ and ‘Bait-and-Switch’). Since our system relies on screenshots, in this work, we limit our analysis to static deceptive patterns, as defined in~\Cref{fig:taxonomy}. The detection of dynamic patterns requires temporal analysis of the webpages, which is out of scope for this work. Our system presents a building block for future systems to detect dynamic patterns by analyzing webpage code, temporal activities, and performing actions.}


\revision{We} filter Gray et al.'s ontology~\cite{gray2023towards} focusing only on \revision{static} deceptive patterns.
 \Cref{fig:taxonomy} shows the filtered taxonomy, comprising four high-level categories and 11 low-level sub-types. The complete taxonomy is provided in~\Cref{app:taxonomy}. This taxonomy includes several high-level categories from Gray et al.~\cite {gray2023towards}. It includes the sub-types from the Brignull et al. taxonomy~\cite{dark_pattern_site}, which map to low-level patterns in Gray et al.'s ontology. These sub-types include textual descriptions of the pattern, which we use to prompt the LLMs, as shown later.  We map our taxonomy to Gray et al.'s ontology~\cite{gray2023towards} in~\Cref{app:mapping}. Finally, the taxonomy excludes a subset of the deceptive patterns that cannot be detected through screenshots, \revision{i.e., the dynamic ones.}


\subsection{Detecting Deceptive Patterns}
\label{subsec:detecting-deceptive-patterns}
Researchers developed several mechanisms to detect and measure the prevalence of online deceptive patterns~\cite{dark_pattern_site,gray2018dark,mathur2019dark,Insite2023,soe2020circumvention,adorna2024developing,chen2023unveiling,raju2022smart,mansur2023aidui}. Curley et al. categorized earlier manual, automated, or semi-automated detection mechanisms~\cite{curley2021design}.  

\subsubsection{Human-based Manual Annotations}
Early efforts to identify online deceptive patterns relied on manual exploration. Brignull et al.~\cite{dark_pattern_site} manually explored the web to compile a ``Hall of Shame'': a list of websites with deceptive patterns. Gray et al.~\cite{gray2018dark} expanded this corpus by performing keyword searches on social media for posts highlighting websites with deceptive patterns, which were then manually validated. While highly accurate, this methodology is limited to domain experts and lacks scalability due to manual validation requirements.
To expedite the manual exploration process, Mathur et al.~\cite{mathur2019dark} proposed a clustering-based pipeline to group similar websites based on their text content, which is then manually inspected. Although this process accelerates data collection, it still lacks scalability.

\subsubsection{Probabilistic Text \& Image Models}
Attempts to automate the manual inspection process include Tung et al.'s \textit{Naive Bayes} classifier~\cite{Insite2023}, Soe et al.'s gradient-boosted tree to flag texts showcasing deceptive patterns~\cite{soe2020circumvention}, and Adorna et al.'s combined \textit{Naive Bayes} classifier and \texttt{VGG-19} model to identify deceptive designs in cookie notices~\cite{adorna2024developing}. However, these works are often domain-specific and cannot readily generalize to new domains. Additionally, most works rely solely on text-based classifiers or heuristics, limiting their ability to detect visual deceptive patterns, such as those based on colors or trick wording.

\subsubsection{Heuristic Based Methods}
Recent works by Chen et al., Mansur et al., and Raju et al. \cite{chen2023unveiling,mansur2023aidui, raju2022smart} focus on automated detection of deceptive patterns in mobile apps. Chen et al.~\cite{chen2023unveiling} and Mansur et al.~\cite{mansur2023aidui} used predefined heuristics, limiting detection to simpler deceptive patterns like disguised ads. Prior works used the \texttt{HTML} code of websites to identify and detect deceptive patterns. For example, Raju et al.~\cite{raju2022smart} employed rule-based source code analysis to detect patterns like ads, forced action, and nagging ~\cite{raju2022smart}. However, due to the versatile and open nature of \texttt{HTML}, this task proves to be very challenging. While \texttt{HTML} implementation and code vary significantly across webpages, and even across different accesses of the same page, screenshots and text remain more consistent. As a result, recent approaches have moved completely towards text or screenshots of websites to detect and identify deceptive designs. 


\subsubsection{Classical ML Models}
In the broader domain of Privacy, Safety, and Security (PSS), researchers have developed tools to help users navigate specific deceptive designs on websites. For instance, works by Khandelwal et al.~\cite{khandelwal_prisec_2021, khandelwal_cookie} enable users to find and adjust privacy settings and automatically disable non-essential cookies. Similarly, OptOutEasy by Kumar et al.~\cite{bannihatti_kumar_finding_2020} automatically finds opt-out links from privacy policies and surfaces them to users. These domain-specific approaches do not readily apply to deceptive pattern detection as they require retraining for each deceptive pattern.

\subsubsection{Large Language Models (VLLMs)}

\revision{Prior work have explored using LLMs to detect deceptive patterns. Sazid et al.~\cite{sazid2023automated} used GPT-3.5-Turbo to detect deceptive text with 92.57\% accuracy, athough their method is limited to singular text lines, only 7 types of patterns, and ignores any surrounding visual or textual context. Similarly, Schäfer et al.~\cite{schafer2025don} utilized GPT-4o to remove deceptive elements from synthetic HTML, reducing manipulativeness in 91\% of cases. This approach, however, is constrained due to the versatile and non-standardized nature of HTML~\cite{tan2025htmlrag} and the challenge of fitting inflated, real-world website source code into an LLM's context window. }

More recently, a concurrent work with ours, Shi et al.~\cite{50Shades}, introduces \textit{DPGuard} to automatically detect deceptive patterns from screenshots of mobile apps and websites by directly prompting VLLMs. We show later in~\Cref{sec:system-overview} that VLLMs perform poorly in detecting deceptive patterns from screenshots and are often prone~to hallucinations and false positives. Additionally, \textit{DPGuard} only detects the presence of these deceptive patterns without any positional reference. Our work identifies deceptive patterns with significantly higher accuracy and extracts their positions, enabling us to show users exactly where these patterns occur.

\section{System Overview}
\label{sec:system-overview}

\begin{figure*}[t]
    \centering
  \includegraphics[width=\textwidth]{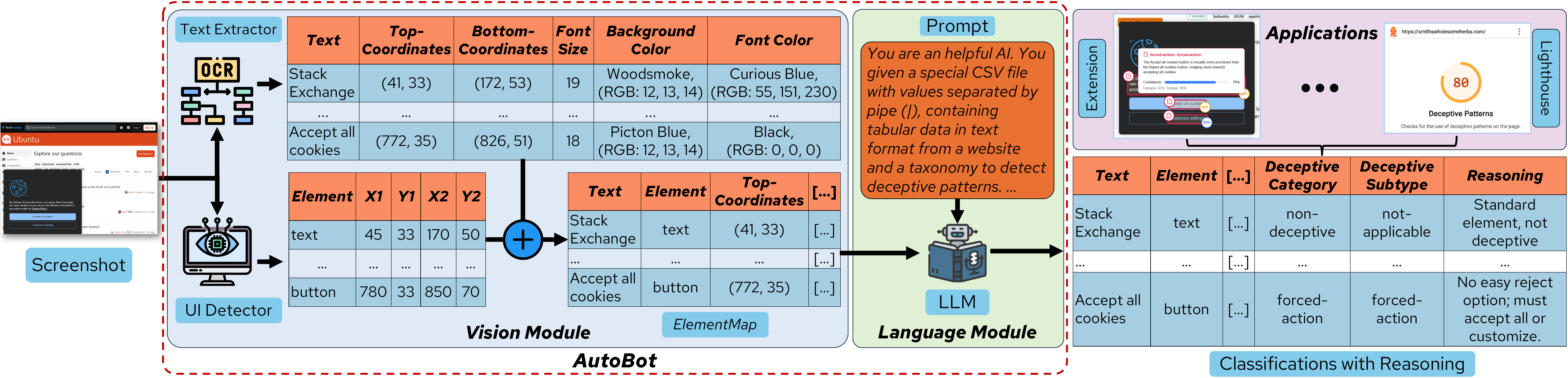}
    \caption{
    Overview of \name's working process. \name takes a screenshot through a multi-stage framework consisting of the Vision Module and the Language Module, and returns the Deceptive Pattern classification, Subtype, and reasoning for each element in the screenshot. The results are then used by applications such as browser extensions and Lighthouse.}
    \label{fig:overview}
\end{figure*}

This work presents \name, an end-to-end framework that identifies the deceptive patterns and extracts their positions on a webpage. After receiving a screenshot of a webpage, it identifies UI elements on the page and feeds the identified elements along with the associated text to an LLM. The output is a mapping of each element to a corresponding deceptive pattern from \Cref{subsec:taxonomy}.

\paragraph{Why Screenshots?}
Prior works analyzing websites~\cite{khandelwal_prisec_2021, khandelwal_cookie} have primarily relied on \texttt{HTML} analysis. This approach faces significant challenges due to the dynamic nature of websites. Websites increasingly employ \textit{JavaScript} frameworks that modify the Document Object Model (DOM) on the fly, rendering static \texttt{HTML} analysis insufficient. Furthermore, the diversity in the coding practices and obfuscation techniques provide additional challenges in \texttt{HTML}-based analyses. To address these challenges, \name adopts a novel approach focusing on the invariant aspect of websites: \textit{the user experience}. By leveraging the visual signals and associated text, \name models how users perceive and interact with websites. This approach offers several advantages: (1) It is resilient to change in underlying technologies as it captures the actual rendered content. (2) It allows us to analyze the same information that the user encounters, providing a more accurate representation of the potential deceptive patterns.

\paragraph{Vision Large Language Models (VLLMs)}
Recent advances in VLLMs offer a venue for analyzing screenshots and highlighting patterns with proper prompting. To this end, we empirically evaluate the effectiveness of \texttt{GPT4.5}~\cite{achiam2023gpt} and \texttt{Gemini 2.5 Pro}~\cite{GoogleGemini} in detecting web deceptive patterns. Our experiments showed that while these models can detect deceptive patterns, but they often hallucinate and give false positive answers. As shown in \Cref{fig:gemini_chat}, both \texttt{Gemini 2.5 Pro} and \texttt{GPT-4.5} struggle to identify the deceptive patterns in screenshots.
A recent work in this domain by Shi et al.~\cite{50Shades}, \textit{DPGuard}, uses these models directly to detect, not localize, deceptive patterns. Consistent with Shi et al.'s evaluation, we find that \textit{DPGuard} faces performance issues, struggling to generalize over a variety of websites (refer to~\Cref{sec:evaluation}), achieving only a macro score of \texttt{0.3452} in detecting deceptive patterns on websites. 

In addition, VLLMs frequently struggle with the precise location of elements in an input image~\cite{chen2023shikra,you2023ferret,li2025towards}, and cannot be used to localize deceptive patterns out of the box. However, when given bounding box coordinates, these models can effectively reason about the spatial arrangements of objects~\cite{shiri2024empirical}. 

\paragraph{Modules}
 \name leverages the above insights to automatically detect and localize deceptive patterns on websites, as shown in \Cref{fig:overview}. It breaks the localization problem and deceptive pattern detection into two tasks: vision and language. This breakdown allows \name to independently leverage vision techniques for the precise localization of elements and powerful language models for the accurate classification of patterns.\smallskip
 
\begin{enumerate}
    \item \textbf{Vision Module:} The \textit{Vision Module} maps a screenshot of a webpage to a table of elements as shown in~\Cref{fig:overview}. We refer to this tabular representation as \textit{\textmap}. The \textit{\textmap} contains the text associated with the element along with its other features: element type, bounding box coordinates, font size, background color, and font color. This module addresses the issues of high false positive rates and localization by parsing the screenshot of a webpage. 
    
    \item \textbf{Language Module:} The \textit{Language Module} (\Cref{sec:language}) takes the \textmap as input and maps each element to a deceptive pattern from the taxonomy in~\Cref{subsec:taxonomy}. This module reasons about each element considering its spatial context and visual features. We explore different instantiations of this module with different trade-offs in terms of cost, need for training, and accuracy.
\end{enumerate}

\section{Vision Module}
\label{sec:vision}


The \textit{Vision Module} generates an \textit{\textmap} from a screenshot of the website through the following three steps, as shown in \Cref{fig:overview}.
\begin{enumerate}
    \item \textbf{Text Extraction:} Extract the text, the bounding boxes of the text, and the associated features from the screenshot.
    \item \textbf{Web-UI Element Detection:} Localize UI elements, extract their bounding boxes, and identify their types from the screenshot.
    \item \textbf{{\textmap} Generation:} Merge the results of the above steps into a tabular representation of the website to generate an \textmap.
\end{enumerate}

\subsection{Text Extraction}

The Text Extraction step starts by performing Optical Character Recognition (OCR) on the website screenshot. We employ the Google Vision OCR API~\footnote{\label{google-ocr}\url{https://cloud.google.com/vision/docs/ocr}} as it has high accuracy in extracting text from images of varying scales and resolutions. The API returns blocks of detected text. We concatenate these blocks based on proximity to form coherent text blocks. These blocks are a list of bounding boxes with their respective text content. Then, we retrieve the font size, font color, and background color of each bounding box, as illustrated in~\Cref{fig:overview}. We calculate font size by subtracting the bottom y-coordinate from the top y-coordinate of the bounding box. To determine color information, we utilize the \textit{extcolors}~\cite{extcolors} package, extracting the most prominent color (background) and the second most prominent color (font).

This approach addresses a key limitation identified by Soe et al.~\cite{soe2020circumvention} in the previous ML methods: the lack of representation of the UI richness that a user perceives, such as text placement and contrast between the font and background colors. Our approach captures these detailed text features, along with the relative location of text elements on the website, providing a comprehensive representation of the textual content experienced by users.


\subsection{Web-UI Element Detection}

The Web-UI Element Detection step uses the same screenshot to identify and localize these 7 \textit{Web-UI Elements}: buttons, checkboxes (\texttt{\faCheckSquare}, \texttt{\faSquare}), radio buttons (\texttt{\faDotCircle}, \texttt{\faCircleNotch}), and toggle switches (\texttt{\faToggleOn}, \texttt{\faToggleOff}). This step provides context to the extracted text and enables a more comprehensive understanding of the webpage's structure. For instance, distinguishing between a clickable button and a static text block can be significant, as a seemingly simple text might be a deceptive call to action when recognized as a button. Similarly, identified checkbox states (checked or unchecked) can reveal pre-selected options that users might overlook.

In the following, we describe how we survey existing Web-UI detection methods and the underlying datasets. As we find these methods and datasets to not be appropriate for the deceptive patterns detection task, we describe how we train a real-time \textit{Web-UI Element Detector} using \texttt{YOLOv10} on a custom dataset.


\paragraph{Limitations of Existing Web-UI Detection Methods}
While Web-UI Element Detection is a well-studied problem, the existing approaches face several limitations in our deceptive pattern detection task. Detectors like \textit{Ominparser 2}~\cite{lu2024omniparser}, \textit{Ferret UI 2}~\cite{li2024ferret}, \textit{UEID}~\cite{xie2020uied}, and \textit{Element Detector}~\cite{wu2023webui}  do not distinguish between checked and unchecked states. Detectors like \textit{UISketch}~\cite{sermuga2021uisketch} report low performance on real-world websites. \textit{ScreenRecognition}~\cite{zhang2021screen} shows promise for this step but was trained on mobile screenshots and is closed-source, preventing its use or testing on websites.

\paragraph{Limitations of Existing Web-UI Datasets}
To improve the detector performance, we explored modifying existing datasets used to train them. \textit{Omniparser 2} uses a manually annotated dataset of popular websites, but the dataset is not public~\cite{lu2024omniparser}. \textit{UEID}~\cite{xie2020uied} was trained on the \textit{RICO Dataset}~\cite{rico}, which contains manually annotated mobile app UIs. We could not modify this dataset to train a detector for webpages. \textit{Ferret UI 2}~\cite{li2024ferret} and \textit{Element Detector}~\cite{wu2023webui} use \textit{WebUI}~\cite{wu2023webui}, a popular framework with 400k websites which is publicly available, but the automatically computed labels from accessibility trees\footnote{Accessibility trees are generated using aria labels which developers have to optionally add to a website.} are noisy and miss a lot of elements on a given website~\cite{wu2023webui,wu_never-ending_2023}. \textit{UISketch} contains hand-drawn images to identify \textit{Web-UI Elements} based on their sketches and does not generalize well to real-world websites~\cite{sermuga2021uisketch}.

\subsubsection{Dataset Curation}
\begin{figure}[t]
    \centering
  \includegraphics[width=\columnwidth]{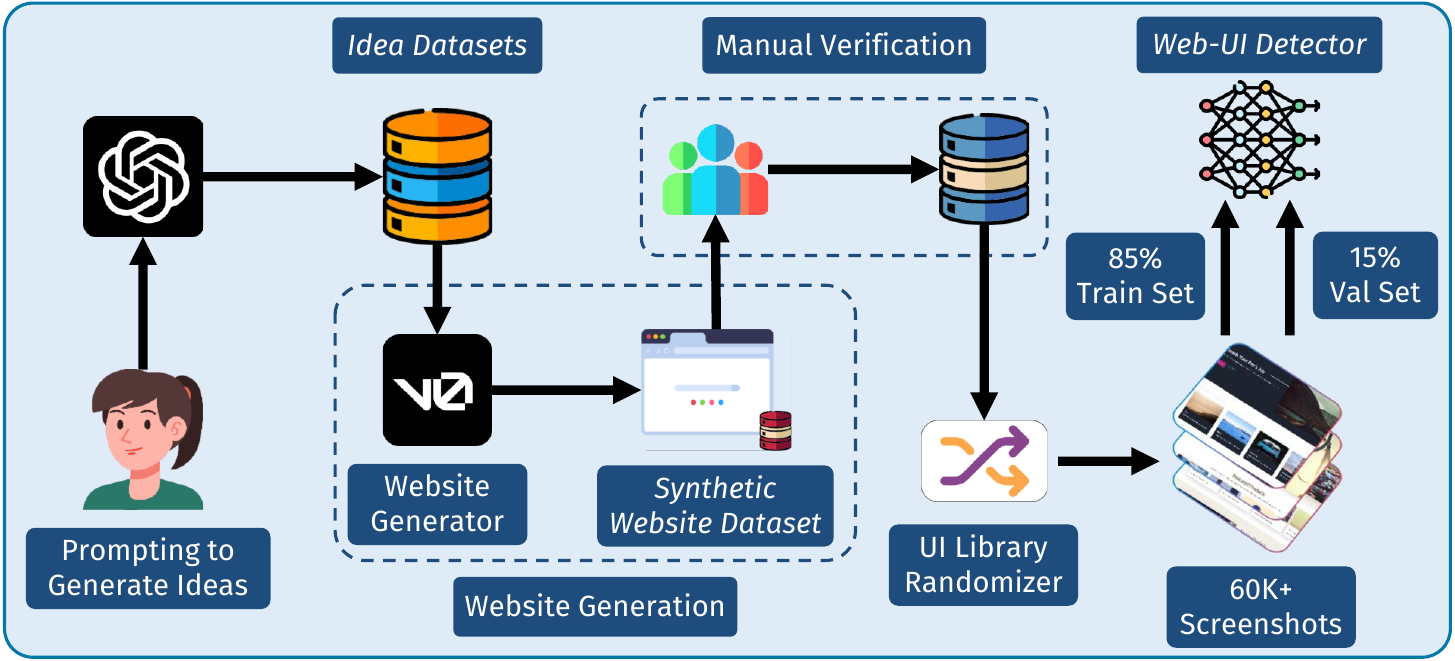}
    \caption{Pipeline of Generating Web-UI Element Dataset to train YOLOv10. \revision{We used GPT-4 to generate 2.5K ideas (\textit{Idea Datasets}), which were then processed by \texttt{v0} to create 7.5K websites (\textit{Synthetic Website Dataset}). After manually verifying these sites for rendering errors and randomizing their UI library, we capture over 60K screenshots to train our YOLOv10 model.}}
    \label{fig:dataset-creation}
\end{figure}

We develop a novel approach to create a diverse and representative dataset to address the lack of suitable \textit{Web-UI Element Dataset} for training our \textit{Web-UI Element Detector}, as shown in \Cref{fig:dataset-creation}. Instead of relying on manual scraping and labeling~\cite{zhang2021screen, sermuga2021uisketch, ueid_xie, lu2024omniparser}, we leverage AI-powered tools to generate a synthetic dataset that reflects the current web landscape. Our approach allows for precise control over element positioning and labeling, which allows us to scale the dataset generation.



We generated 2.5K ideas for diverse websites using \texttt{GPT-4}~\cite{achiam2023gpt}. We passed these ideas to \texttt{v0}\footnote{\url{https://v0.dev/}} (refer to~\Cref{app:website-generation} for examples), an AI-based website generator, to generate three websites per idea. \texttt{v0} uses \texttt{shadcn}~\cite{shadcn}, a customizable UI-Library, which allows us to get bounding boxes and the state (checked/unchecked) of every UI element. Three members of the research team manually verified the 7.5K websites generated by \texttt{v0}. With a failure rate of less than 2\%, this process was relatively fast, as errors in websites are detected by the compiler. These errors consisted mainly of typos and missing libraries that the researchers manually fixed.
We then rendered each \texttt{v0} generated website (7.5K) using randomized components from 6 popular UI-libraries, like Material UI and Bootstrap\footnote{\url{https://mui.com/material-ui/}, \url{https://getbootstrap.com/}}. This process resulted in 62K screenshots, where each screenshot had a bounding box and a label for each UI element. Recall that we have 7 labels for the UI elements. We refer to this dataset as the \textit{Web-UI Elements Dataset}.

\subsubsection{Training Web-UI Element Detector}
We use the \textit{Web-UI Elements Dataset} to train an ensemble of \texttt{YOLOv10} models. In particular, we randomly divide the dataset into a training set consisting of 85\% of the images and a validation set consisting of the remaining 15\%. We present the performance of the trained ensemble in \Cref{subsec:eval:vision} on real-world websites.


\paragraph{Why YOLOv10?}
We adopt the \texttt{YOLOv10} model (You Only Look Once (YOLO) architecture~\cite{wang2024yolov10}, a real-time object detector for recognizing UI elements from a screenshot. We chose \texttt{YOLOv10} over Convolution Neural Networks (CNNs)~\cite{khandelwal_prisec_2021, mansur2023aidui, chen2023unveiling} and VLLMs like \texttt{Molmo}~\cite{molmo}. CNN-based detectors, such as \textit{Faster R-CNN}~\cite{ren2015faster}, provide predictions with high accuracy, but require considerable computational power and time~\cite{redmon2018yolov3, ren2015faster}. VLLMs, despite their strong capabilities in understanding image context, demonstrate significant limitations in image classification tasks~\cite{zhang2024visually}, specifically object detection tasks~\cite{zang2025contextual}. Evaluating \texttt{Molmo}~\cite{molmo} on detecting \textit{Web-UI Elements} yielded poor results, as shown in~\Cref{tab:yolo_molmo}. \texttt{YOLO} models are comparatively lightweight, around 40MB, allowing for various deployment options without requiring extensive compute resources.






\paragraph{Ensemble of YOLOv10}
During training, we observed that \texttt{YOLOv10} models could not distinguish between the 7 labels (first column of \Cref{tab:yolo_molmo}) accurately. We attribute the reason to the labels being visually similar, such as a checked switch and a checked radio button. As such, we trained three \texttt{YOLOv10} models, each focused on distinguishing between 2-3 different elements. The first model labeled button and \texttt{\faDotCircle}; the second labeled \texttt{\faToggleOn}, \texttt{\faCheckSquare}, and \texttt{\faSquare}; and the third labeled \texttt{\faToggleOff} and \texttt{\faCircleNotch}. We found that each model performed much better than one trying to handle all the classes at once. The same observation has been made in literature before for YOLO-based object detection~\cite{mohankumar_benchmark_nodate, walambe_lightweight_2021, pham_optimizing_2024}. We combined the outputs of the three models by simply performing a union over the detected UI elements. In case of overlap, we took the label with the higher confidence classification. We refer to this detection method as the \texttt{YOLOv10 Ensemble}.




\subsection{\textmap Generation}
The \textmap Generation step merges the \textit{Web-UI Elements} from the \textit{Web-UI Element Detector} with the text blocks from the \textit{Text Extraction} step. In particular, it iterates over each detected \textit{Web-UI Element} and applies spatial heuristics depending on the element type to find the most likely text block corresponding to the element. For example, buttons are matched based on overlap, while checkboxes and radio buttons are paired with nearby text. The closest matching text block is then relabeled with the element type. This labeling results in an \textit{\textmap}, where each row contains an element label, the text, the bounding box coordinates, the font size, the background color, and the font color. The \textit{Language Module} uses the \textmap to detect the deceptive patterns on a page.

\subsection{Vision Module Evaluation}
We create a dataset to evaluate the real-world performance of the \texttt{YOLOv10 ensemble}. 

\subsubsection{Vision Dataset}
We curate a labeled dataset of UI elements from the deceptive pattern websites dataset of Mathur et al.~\cite{mathur2019dark}. We manually annotated over 1.5K website screenshots using Label Studio~\cite{Label_Studio}. \revision{In particular, one author manually annotated each screenshot by drawing bounding boxes around each UI element and assigning it a type. The type is one the 7 \textit{Web-UI Elements}: buttons, checkboxes (\texttt{\faCheckSquare}, \texttt{\faSquare}), radio buttons (\texttt{\faDotCircle}, \texttt{\faCircleNotch}), and toggle switches (\texttt{\faToggleOn}, \texttt{\faToggleOff}). Another author independently verified the annotations. Both authors then discussed and resolved the conflicts in annotations.} We refer to this dataset as the \textit{Real-UI Dataset}.

\subsubsection{Vision Module Evaluation}
\label{subsec:eval:vision}
We measure the accuracy of the \texttt{YOLOv10 ensemble} of models on the \textit{Real-UI Dataset}. We report our results using the IoU metric, which measures the overlap between the predicted bounding box and the ground truth box by dividing the area of their intersection by the area of their union. A higher IoU indicates better localization accuracy, and we consider a detection to be correct if the IoU exceeds 0.5 and the model confidence exceeds 0.3. We choose a lower IoU threshold to account for the distribution shift between training on a synthetically labeled dataset and a human-annotated one. The bounding boxes from both will be different. Using a high IoU threshold would result in more false negatives, which would affect the subsequent steps in \name's pipeline. 

\begin{table}[ht]
\caption{Performance of the \textit{YOLOv10 Ensemble} and \textit{Molmo}\revision{~\cite{molmo}} on our \textit{Real-UI Dataset}}
\centering
\resizebox{\columnwidth}{!}{
\begin{tabular}{ l c c c c c c c}
\toprule
\multirow{2}{*}{\textbf{Class}} & \multicolumn{2}{c}{\textbf{Precision}} & \multicolumn{2}{c }{\textbf{Recall}} & \multicolumn{2}{c}{\textbf{F1-Score}} & \multirow{2}{*}{\textbf{\# Elements}} \\ \cmidrule{2-7}
 & \textbf{YOLO} & \textbf{Molmo} & \textbf{YOLO} & \textbf{Molmo} & \textbf{YOLO} & \textbf{Molmo} & \\
\midrule
\rowcolor{aliceblue}
\textbf{button} & \textbf{0.94} & 0.87 & \textbf{0.88} & 0.41 & \textbf{0.91} & 0.55 & 5226 \\
\textbf{\faCheckSquare} & 0.85 & \textbf{0.97} & \textbf{0.95} & 0.47 & \textbf{0.89} & 0.63 & 113 \\
\rowcolor{aliceblue}
\textbf{\faSquare} & \textbf{0.98} & 0.92 & \textbf{0.76} & 0.44 & \textbf{0.86} & 0.60 & 246 \\
\textbf{\faDotCircle} & 0.85 & \textbf{0.86} & \textbf{0.93} & 0.29 & \textbf{0.89} & 0.43 & 76 \\
\rowcolor{aliceblue}
\textbf{\faCircleNotch} & \textbf{0.86} & 0.84 & \textbf{0.89} & 0.35 & \textbf{0.87} & 0.49 & 132 \\
\textbf{\faToggleOn} & \textbf{0.96} & \textbf{0.96} & \textbf{0.98} & 0.42 & \textbf{0.97} & 0.59 & 52 \\
\rowcolor{aliceblue}
\textbf{\faToggleOff} & 0.91 & \textbf{0.93} & \textbf{0.94} & 0.47 & \textbf{0.93} & 0.63 & 34 \\ \midrule
\textbf{Total} & \textbf{0.91} & 0.90 & \textbf{0.91} & 0.42 & \textbf{0.91} & 0.57 & 5879 \\
\bottomrule
\end{tabular}
}
\label{tab:yolo_molmo}
\end{table}

\sloppy
\Cref{tab:yolo_molmo} shows the F1-scores of our \texttt{YOLOv10 Ensemble} and \texttt{Molmo}~\cite{molmo} for each of the 7 classes. We observe that the ensemble outperforms \texttt{Molmo} for all the UI elements. Note that few-shot prompting of local VLLMs is not possible for image inputs.  The ensemble exhibits a relatively lower performance on the unchecked radio button and unchecked check box, because they look visually similar to the other UI elements.

\section{Language Module}
\label{sec:language}

\begin{figure}[t!]
    \centering
  \includegraphics[width=\columnwidth]{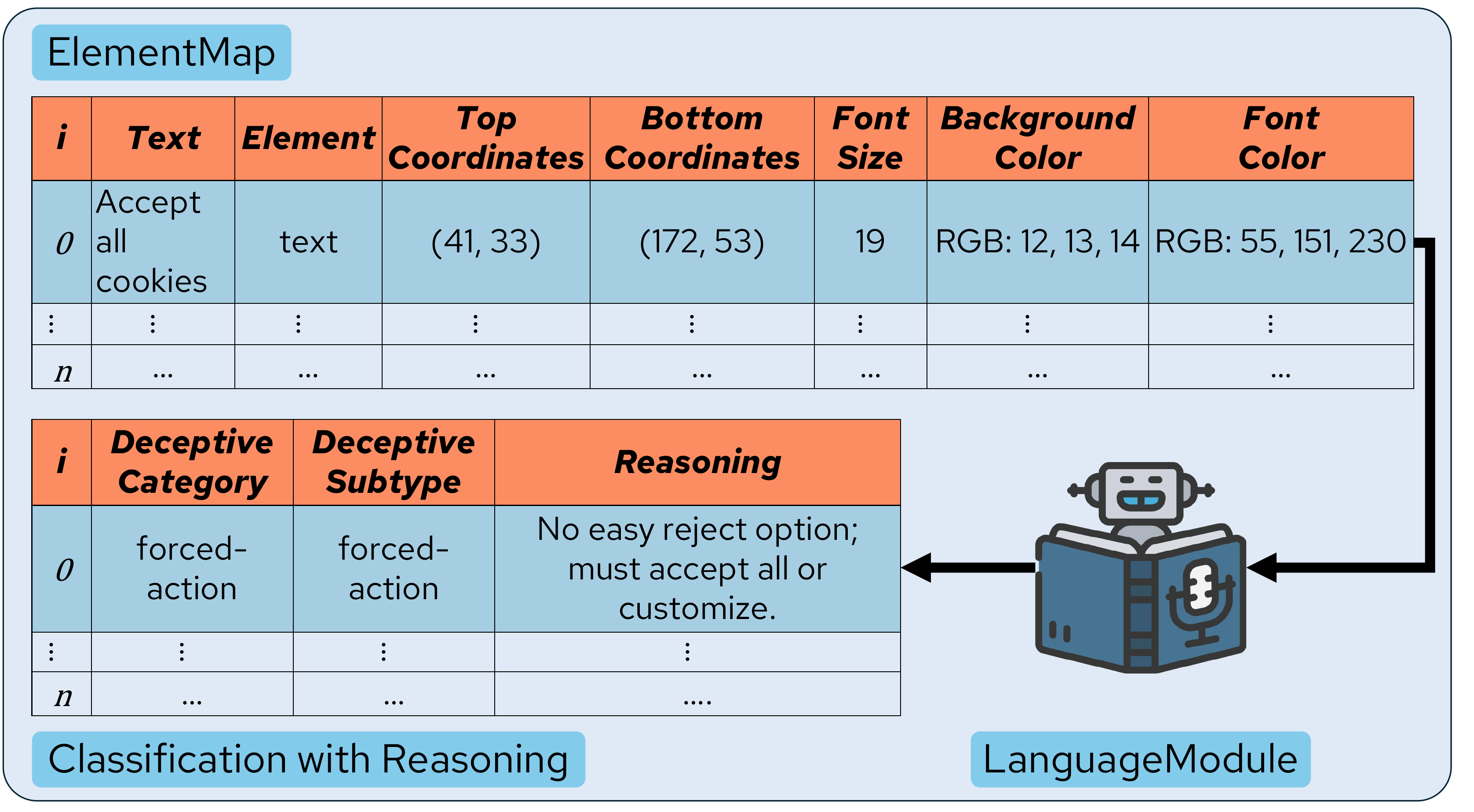}
    \caption{The input and output structure of our language module. The input ElementMap consists of key features of a web element, and the output contains a deceptive category, subtype, and reasoning of classification.}
    \label{fig:language-module}
\end{figure}

The \textit{Language Module} assigns a deceptive pattern (from the taxonomy in~\Cref{subsec:taxonomy}) to each element present in an \textmap. This is the input/output structure of the language module as shown in~\Cref{fig:language-module}.



\subsection{Possible Solutions}
\label{sub:possible-solutions}

Prior works have shown that LLMs~\cite{nayak2024experimental} are suitable for reasoning tasks similar to our task. 
These models fall into three categories: 1) large Language Models (LMs) like \texttt{Gemini} and \texttt{GPT-4}, 2) small LMs such as \texttt{Gemma} and \texttt{Qwen}, and 3) very small LMs like \texttt{T5}. These models have varying capabilities and trade-offs, as described in~\Cref{tab:model-comparison}.

\begin{table}[htbp]
\centering
\caption{Comparison of Language Models: \texttt{Gemini} vs \texttt{Qwen2.5} vs \texttt{T5}}
\resizebox{\columnwidth}{!}{
\begin{tabular}{@{}l@{\hspace{0.2cm}}l@{\hspace{0.2cm}}l@{\hspace{0.2cm}}l@{}}
\toprule
\textbf{Feature} & \textbf{Large LM} & \textbf{Small LM} & \textbf{Very Small LM} \\
\midrule
\rowcolor{aliceblue}
Size (parameters) & Large (>200B) & Medium (1.5B) & Very Small (700M) \\
Context Window & 1M & 128K & $<$1K \\
\rowcolor{aliceblue}
Deployment & Cloud API only & Can be run locally & Can be run locally \\
Required Memory & N/A & $\sim$4.5 GB & $\sim$700MB \\
\rowcolor{aliceblue}
License & Proprietary & Open Source & Open Source \\
Latency & Higher & Medium & Very low \\
\rowcolor{aliceblue}
Cost & High & Free & Free \\
Data Privacy & Data leaves device & Data stays on device & Data stays on device \\
\bottomrule
\end{tabular}
}
\label{tab:model-comparison}
\end{table}

\subsubsection{Large Language Models (LLMs):}
LLMs are very effective at performing a wide range of tasks~\cite{cui2025curie, agarwal2024llm, shool2025systematic} and are able to closely follow user instructions~\cite{ouyang2022training,zhou2023instruction}. 
However, as shown in~\Cref{tab:model-comparison}, using these LLMs is cost-prohibitive and has potential data privacy concerns~\cite{api_security}.


\subsubsection{Small Language Models (SLMs):}
\label{subsec:qwen}
Unlike LLMs that follow complex instructions when performing a task, SLMs often struggle to do so. This limitation of SLMs can be overcome by fine-tuning them on specific tasks, improving their performance to match that of LLMs~\cite{deepscaler2025}. Moreover, as shown in~\Cref{tab:model-comparison}, SLMs have two key advantages over LLMs: 1) being compute efficient, they can be deployed locally across a wide range of platforms, and 2) they alleviate any data privacy concern associated with API based LLMs.



\subsubsection{Very Small Language Models (vSLMs):}
\label{subsec:t5}
SLMs, while compute-efficient, are not the best solution to use in an extremely resource-constrained platform, such as in-browser or on devices with no specialized GPU. In such environments, we can leverage vSLMs like \texttt{Flan-T5}~\cite{FlanT5}. Models like \texttt{Flan-T5} need to be finetuned or distilled from LLMs for specific tasks to achieve high accuracy~\cite{hsieh2023distilling}. We show the detailed distillation steps we performed in \Cref{subsec:distil-t5}.


In summary, we observe that large LMs, such as \texttt{Gemini}, are considerably effective in detecting deceptive patterns from an \textmap. However, as described in \Cref{tab:model-comparison}, utilizing such large and closed-source models presents challenges like high usage cost, considerable latency, and potential data-privacy concerns as the \textmap is sent to an external service. Smaller LMs such as \texttt{Qwen} and \texttt{T5} address these challenges and have been shown to perform well on such specific tasks after finetuning~\cite{deepscaler2025,hsieh2023distilling}. 

\subsection{Our Solution}
To combine the strengths of large and small LMs, we adopt a distillation approach where we use large LMs as teachers and small LMs as students to complete our task. Specifically, we create a synthetic dataset of deceptive pattern classification from \textmap using \texttt{Gemini}. We then use this dataset to distill smaller student models, i.e., \texttt{Qwen} and \texttt{T5}. As such, \name comprises three language models to detect deceptive patterns, each presenting different trade-offs as described in \Cref{tab:model-comparison}.

\subsubsection{Prompting the LLM}

We utilize proven techniques like Chain-of-Thought (CoT)~\cite{wei2022chain}, few-shot prompting~\cite{brown2020language}, and prompting the model to reason about its classification~\cite{openai_learning_reason_2024,guo2025deepseek,nayak2024experimental} to help the model understand the task and identify deceptive patterns. Specifically, our system prompt, $P_\text{system}$\footnote{\label{fn:sys_prompt}\revision{Please find the system prompt under \textit{Files} here: \url{https://osf.io/tha2d/?view_only=4fd2116fa2e94c99857679eddfea5937}}}, provides the LLM with a plan on how to detect deceptive patterns and instructs the LLM to generate three columns — \textit{Deceptive Category}, \textit{Deceptive Subtype}, and \textit{Reasoning} — for each element in the \textmap, as shown in~\Cref{fig:language-module}. 
This prompt allows us to not only generate precise labels and the associated reasoning but also minimize hallucinations. An added benefit of generating the classification with `Reasoning' is that we can use these `rationales' to further train smaller LMs. This
\textit{Distilling Step-by-Step} methodology has been shown to be very effective by Hsieh et al.~\cite{hsieh2023distilling}.

During our initial evaluation of \texttt{Gemini 1.5 Pro}, using $P_\text{system}$, we observed that the model could identify all deceptive patterns in the \textmap, but would often mis-classify `non-deceptive' patterns as deceptive, resulting in a high false positive rate and a low precision score.
To address this problem, we utilized the \texttt{Gemini 2.0-Flash-Thinking} reasoning model. Reasoning models have shown promising results in being able to reason about a task~\cite{deepscaler2025,openai_learning_reason_2024}. 
We prompted the LLM to re-evaluate the elements identified as deceptive and correct misclassifications\revision{\footref{fn:sys_prompt}}.
Next, we utilized the Gemini 2.0-Flash-Thinking model to re-verify the labels on every website with one or more elements classified as deceptive.
This additional step in generating the labels, significantly reduced the number of false positives we observed.
We note that we do not use this model as our base model because of its limited availability, making it infeasible to be used at scale. We show that this approach results in high precision and recall in~\Cref{subsec:eval:lang}.



\subsubsection{Creating Distillation Dataset}
\label{subsub:d3-dataset}
We create \distill by scraping and analyzing 11K websites from the Tranco list~\cite{pochat2018tranco}. We filter out non-English websites using the \texttt{langdetect} library~\cite{langdetect} and adult websites using the \texttt{NudeNet} package~\cite{nudenet}, leaving us with 6,626 websites.
For these websites, a significant portion was non-deceptive or had very few deceptive patterns. 
To increase websites with deceptive patterns, we opted to look at e-commerce websites, as these websites tend to have deceptive patterns~\cite{mathur2019dark}. As such, we analyze an additional 4,492 featured websites from \textit{Shopify Partners Directory}\footnote{\url{https://www.shopify.com/partners/directory}}.

Overall, we have 11,118 websites in our dataset. We run \name on these websites, using the \texttt{Gemini 1.5 Pro} and \texttt{Gemini 2.0 Flash-Thinking} models. To reduce randomness and have deterministic outputs, we limit the $temperature=0$ and $top_p=0.1$~\cite{renze2024effect}. We incorporate the final classification and reasoning produced by \name in \distill. The distribution of samples (a sample is defined as a single row of the \textmap) is shown in~\Cref{tab:d_distill}.

\begin{table}[h]
\caption{Distribution of Samples in \distill.}
\centering
\resizebox{0.9\columnwidth}{!}{
\begin{tabular}{l l c}
\toprule
\textbf{Category} & \textbf{Subtype} & \textbf{\# Samples} \\
\midrule
\textbf{\revision{Non Deceptive}}  & \textit{\revision{Not Applicable}}  & 160934 \\
\cmidrule(lr){1-3}
\textbf{\revision{Forced Action}}  & \textit{\revision{Forced Action}}  & 3403 \\
\cmidrule(lr){1-3}
\multirow{3}{*}{\centering \textbf{\revision{Interface Interference}}} & \textit{\revision{Nudge}} & 1335 \\
& \textit{\revision{Fake Scarcity / Fake Urgency}} & 933 \\
& \textit{\revision{Confirmshaming}} & 428 \\
\cmidrule(lr){1-3}
\multirow{2}{*}{\centering \textbf{\revision{Obstruction}}} & \textit{\revision{Visual Interference}} & 689 \\
& \textit{\revision{Pre-Selection}} & 234 \\
\cmidrule(lr){1-3}
\multirow{3}{*}{\centering \textbf{\revision{Sneaking}}} & \textit{\revision{Trick Wording}} & 904 \\
& \textit{\revision{Hidden Costs}} & 233 \\
& \textit{\revision{Hidden Subscription}} & 3495 \\
& \textit{\revision{Disguised Ads}} & 5980 \\
\bottomrule
\end{tabular}
}
\label{tab:d_distill}
\end{table}

\subsubsection{Distilling Small Language Models (Qwen2.5-1.5B)}
\label{subsec:distil-qwen}
\sloppy
DeepScaleR~\cite{deepscaler2025} has shown that small language models (SLMs) finetuned for specific tasks are able to mirror the performance of larger models on those same tasks. Using this insight, we distill a \texttt{Qwen2.5-1.5B} model to be able to mimic \texttt{Gemini}'s performance at detecting deceptive patterns. We use the \distill dataset to perform full-finetuning of the \texttt{Qwen2.5-1.5B} model. As shown in~\Cref{tab:d_distill}, the distribution of samples in the distillation dataset is highly imbalanced, with over 92\% of the samples being `non-deceptive'. Training on such an imbalanced dataset may introduce bias towards the majority class~\cite{leevy2018survey}. To mitigate such bias, we perform Random Under Sampling of the `non-deceptive' class, which has shown to improve performance~\cite{elsoud2024under}, to achieve a more balanced distribution of about 55\% ‘non-deceptive’
samples. Our distilled version of \texttt{Qwen2.5-1.5B} was trained on 34.7M tokens on 4 Nvidia-A6000 GPUs for 2 epochs. We present the performance of the trained SLM in~\Cref{subsec:eval:lang}.

\subsubsection{Distilling Very Small Language Models (T5):}
\label{subsec:distil-t5}
To distill the knowledge from LLMs into very small language models like \texttt{T5}, we use and improve the paradigm introduced by Hsieh et al.~\cite{hsieh2023distilling}. 
We split \distill into 90\% training and a 10\% validation set grouped by the sites from which the samples were extracted. This ensures that samples from the same site are not present in training and testing sets. 
We formally define the training dataset, $D_\text{train}$, as:
\begin{gather}
    d_i \in D_\text{train} \subset \mathcal{D_\text{distill}}\\
    d_i = (x_i, y_i, z_i, r_i)
\end{gather}

\noindent
where $D_\text{train}$ is a subset of the balanced \distill from~\Cref{subsec:distil-qwen}.
Here, $x_i$ represents the input to classify, $y_i$ represents the deceptive design category, $z_i$ represents the deceptive design subtype, and $r_i$ represents the associated reasoning for the category and subtype. Since, the context window of \texttt{T5} is significantly smaller than that of SLMs and LLMs, for each web element that is to be classified,~$x_i$, we provide its neighboring web elements, in a sliding window fashion: $x_{i-1\rightarrow i-n}\text{ to }x_{i+1\rightarrow i + n}$. For the distillation task, we used $n = 4$.



\paragraph{Baseline Approach}
Based on this initial methodology, we train a \texttt{T5} model, $f$, on a multi-task problem: predict the deceptive design category, subtype as the label, and reasoning as the rationale. We use the task-specific prefix \texttt{[classify]} to generate the label and \texttt{[reason]} for the reasoning.

We define the model and loss functions as:

\begin{gather}
    f(x, t) =
     \begin{cases}
       (\hat{y}_i \oplus \hat{z}_i), & \text{if } t=\texttt{[category]}\\
       \hat{r}_i, & \text{if } t=\texttt{[reason]}\\
     \end{cases}     
\end{gather}

\begin{gather}
\label{loss_baseline}
    \mathcal{L} = \mathcal{L}_\text{label} + \alpha \mathcal{L}_\text{reason}
\end{gather}

Here, $\mathcal{L}_\text{label}$ and $\mathcal{L}_\text{reason}$ are the label prediction loss and reason generation loss, respectively, and are defined as:

\begin{gather}
    \mathcal{L}_\text{label} = \frac{1}{N}\sum_{i=1}^N \ell(f(x_i, t_\text{category}), y_i \oplus z_i) \\
    \mathcal{L}_\text{reason} = \frac{1}{N}\sum_{i=1}^N \ell(f(x_i, t_\text{reason}),r_i),
\end{gather}
 where $\ell$ is the cross entropy loss between the predicted and target tokens, and $\oplus$ is a string concatenation function.

Each web element that is to be classified, $x_i$, is provided alongside its neighboring web elements, $x_{i-1\rightarrow0}\text{ to }x_{i+1\rightarrow N}$. The model's performance is measured as the exact match of its \texttt{[category]} outputs with the ground truth data. An example of the model's sample input and expected output is shown below.

\resizebox{0.9\columnwidth}{!}{
\begin{samplebox}
  \begin{inputbox}
    \textbf{Input:}\quad\textcolor{bracketcolor}{\texttt{[category]}}\texttt{: Line 14,Preferences,checked checkbox,...}\texttt{</s>}\texttt{Line 10,"MAGIC We use cookies to personalise ...}\texttt{</s>}\texttt{Line 11,COMING SOON ,text,...}\texttt{</s>}
  \end{inputbox}
  
  \begin{outputbox}
    \textbf{Output:} \textit{obstruction,pre-selection}
  \end{outputbox}
  
  
  \begin{inputbox}  
    \textbf{Input:}\quad\textcolor{bracketcolor}{\texttt{[reason]}}\texttt{: Line 14,Preferences,checked checkbox,...}\texttt{</s>}\texttt{Line 10,"MAGIC We use cookies to personalise ...}\texttt{</s>}\texttt{Line 11,COMING SOON ,text,...}\texttt{</s>}
  \end{inputbox}
  
  \begin{outputbox}
    \textbf{Output:} \textit{Cookie banner option is pre-selected to indicate users to allow extra cookies.}
  \end{outputbox}
\end{samplebox}
}

After training \texttt{T5} on the \distill dataset for 2 epochs
, we observed the training accuracy to saturate to \textasciitilde $48\%$. Here, accuracy refers to the correct prediction of both category and sub-types.

\paragraph{Our Approach:} To overcome the low accuracy in the baseline approach, we split the labeling task into two separate tasks: category and subtype, introducing an additional ``task prefix'' \texttt{[subtype]} for the new task. We redefine the model and loss function to:
\begin{gather}
    f(x, t) =
     \begin{cases}
       \hat{y}_i, & \text{if } t=\texttt{[category]}\\
       \hat{z}_i, & \text{if } t=\texttt{[subtype]}\\
       \hat{r}_i, & \text{if } t=\texttt{[reason]}\\
     \end{cases}     
\end{gather}
\begin{gather}
\label{loss_ours}
    \mathcal{L} = \alpha (\mathcal{L}_{\mathrm{category}} + \mathcal{L}_{\mathrm{subtype}})  +  (1 - \alpha) \mathcal{L}_{\mathrm{reason}},
\end{gather}
where $\alpha$ is a tuning factor.

By separating \texttt{label} into \texttt{category} and \texttt{subtype} in addition to the \texttt{reason} tasks, the model can learn the relation between the category and the subtype and how they both relate to the reasoning. A sample of the new inputs and expected outputs is shown below.

\resizebox{0.9\columnwidth}{!}{
\begin{samplebox}
\begin{inputbox}
  \textbf{Input:}\quad\textcolor{bracketcolor}{\texttt{[category]}}\texttt{: Line 14,Preferences,checked check-box,...</s>Line 10,"MAGIC We use cookies to personalise ...</s>Line 11,COMING SOON ,text,...</s>}
  \end{inputbox}
\begin{outputbox}
  \textbf{Output:} \textit{obstruction}
  \end{outputbox}
  \begin{inputbox}
  \textbf{Input:}\quad\textcolor{bracketcolor}{\texttt{[subtype]}}\texttt{: Line 14,Preferences,checked checkbox,...</s>Line 10,"MAGIC We use cookies to personalise ...</s>Line 11,COMING SOON ,text,...</s>}
  \end{inputbox}
   \begin{outputbox}
  \textbf{Output:} \textit{pre-selection}
  \end{outputbox}

  \begin{inputbox}
  \textbf{Input:}\quad\textcolor{bracketcolor}{\texttt{[reason]}}\texttt{: Line 14,Preferences,checked checkbox,...</s>Line 10,"MAGIC We use cookies to personalise ...</s>Line 11,COMING SOON ,text,...</s>}
 \end{inputbox}

  \begin{outputbox}
  \textbf{Output:} \textit{Cookie banner option is pre-selected to indicate users to allow extra cookies.}
  \end{outputbox}
\end{samplebox}
}

After training the \texttt{T5} model on the new loss function and tasks using the same ground truth dataset and the same number of epochs, we observe the test accuracy of detecting deceptive patterns saturate to $\sim\!95\%$. We present these results in the following section.

\subsection{Language Module Evaluation}
\label{subsec:eval:lang}
We create a dataset to evaluate the real-world performance of the different models in the language module.

\subsubsection{Language Dataset}
\revision{To evaluate the \textit{Language Module} of our pipeline, we curate a dataset, referred to as \textit{LangEval} dataset. In particular, we randomly choose 200 websites from the \textit{D3 Dataset} (in particular the Mathur et al. portion) described in \Cref{sub:eval:dataset-creation}. Next, for these 200 websites, one of the authors manually annotated the UI element classifications in the \textmap to provide ground truth UI labels. Thus, the \textit{LangEval} dataset contains the manually labeled \textmap of each website associated with the manually labeled deceptive patterns.
}




\subsubsection{Evaluating Language Models}
We evaluate the various language models discussed in this section on the \textit{LangEval} dataset. The performance is shown in \Cref{tab:langeval_no_t5base}. We report the performance of the language models in two ways:
\begin{enumerate}[leftmargin=*]
    \item[1.] \textit{Binary Classification:} This metric assesses models' ability to detect whether a UI element \textit{is deceptive}, regardless of the specific categorization. We use this metric as it provides a performance measure while considering the inherent subjectivity of deceptive pattern classification -- an element classified as `trick-wording' could also be classified as `hidden-cost'. \revision{This classification is provided at the bottom of each evaluation table, with labels: “Deceptive” and “Non-Deceptive”.}
    
    \item[2.] \textit{Class-wise Classification:} This is the \textit{detailed classification} result, across categories and subtypes, between the ground truth data and the model-generated result.
\end{enumerate}

\begin{table}[h]
\captionof{table}{Performance of different models in \textit{LanguageModule} on \textit{LangEval}. \revision{The table has three sections, each showing performance at the category, subtype, and binary levels}}
\centering
\resizebox{\columnwidth}{!}{%
\begin{tabular}{l p{3cm} ccc ccc ccc}
\toprule
& \multirow{2}{*}{\textbf{Pattern Type}} & \multicolumn{3}{c}{\textbf{Precision}} & \multicolumn{3}{c}{\textbf{Recall}} & \multicolumn{3}{c}{\textbf{F1-Score}} \\
\cmidrule(lr){3-5} \cmidrule(lr){6-8} \cmidrule(lr){9-11}
& & \textbf{Gemini} & \textbf{Qwen} & \textbf{T5} & \textbf{Gemini} & \textbf{Qwen} & \textbf{T5} & \textbf{Gemini} & \textbf{Qwen} & \textbf{T5} \\
\midrule
\multirow{5}{*}{\rotatebox{90}{\textbf{Category}}}
& \cellcolor{aliceblue}\textbf{\revision{Non Deceptive}} & \cellcolor{aliceblue}\textbf{1.00} & \cellcolor{aliceblue}0.98 & \cellcolor{aliceblue}\textbf{1.00} & \cellcolor{aliceblue}\textbf{0.99} & \cellcolor{aliceblue}\textbf{0.99} & \cellcolor{aliceblue}0.93 & \cellcolor{aliceblue}\textbf{0.99} & \cellcolor{aliceblue}0.98 & \cellcolor{aliceblue}0.96 \\
& \textbf{\revision{Forced Action}} & \textbf{0.89} & 0.85 & 0.84 & 0.79 & \textbf{0.83} & \textbf{0.86} & 0.84 & 0.84 & \textbf{0.85} \\
& \cellcolor{aliceblue}\textbf{\revision{\begin{tabular}[c]{@{}l@{}}Interface \\ Interference\end{tabular}}} & \cellcolor{aliceblue}\textbf{0.85} & \cellcolor{aliceblue}0.79 & \cellcolor{aliceblue}0.51 & \cellcolor{aliceblue}\textbf{0.85} & \cellcolor{aliceblue}0.55 & \cellcolor{aliceblue}0.69 & \cellcolor{aliceblue}\textbf{0.85} & \cellcolor{aliceblue}0.65 & \cellcolor{aliceblue}0.59 \\
& \textbf{\revision{Obstruction}} & \textbf{0.53} & 0.49 & 0.17 & 0.91 & 0.95 & \textbf{1.00} & \textbf{0.67} & 0.64 & 0.29 \\
& \cellcolor{aliceblue}\textbf{\revision{Sneaking}} & \cellcolor{aliceblue}\textbf{0.87} & \cellcolor{aliceblue}0.73 & \cellcolor{aliceblue}0.50 & \cellcolor{aliceblue}\textbf{0.96} & \cellcolor{aliceblue}0.71 & \cellcolor{aliceblue}0.84 & \cellcolor{aliceblue}\textbf{0.91} & \cellcolor{aliceblue}0.72 & \cellcolor{aliceblue}0.63 \\
\cmidrule(lr){1-11}
\multirow{11}{*}{\rotatebox{90}{\textbf{Subtype}}}
& \textit{\revision{Not Applicable}} & \textbf{1.00} & 0.98 & 0.99 & \textbf{0.99} & \textbf{0.99} & 0.89 & \textbf{0.99} & 0.98 & 0.94 \\
& \cellcolor{aliceblue}\textit{\revision{Confirmshaming}} & \cellcolor{aliceblue}0.80 & \cellcolor{aliceblue}\textbf{0.90} & \cellcolor{aliceblue}0.86 & \cellcolor{aliceblue}0.80 & \cellcolor{aliceblue}0.82 & \cellcolor{aliceblue}\textbf{1.00} & \cellcolor{aliceblue}\textbf{0.80} & \cellcolor{aliceblue}0.86 & \cellcolor{aliceblue}0.92 \\
& \textit{\revision{Disguised Ads}} & \textbf{0.76} & 0.58 & 0.33 & \textbf{0.96} & 0.58 & 0.73 & \textbf{0.85} & 0.58 & 0.46 \\
& \cellcolor{aliceblue}\textit{\revision{\begin{tabular}[c]{@{}l@{}}Fake Scarcity / \\ Fake Urgency\end{tabular}}} & \cellcolor{aliceblue}\textbf{0.96} & \cellcolor{aliceblue}0.91 & \cellcolor{aliceblue}0.61 & \cellcolor{aliceblue}\textbf{0.93} & \cellcolor{aliceblue}0.63 & \cellcolor{aliceblue}0.76 & \cellcolor{aliceblue}\textbf{0.95} & \cellcolor{aliceblue}0.75 & \cellcolor{aliceblue}0.68 \\
& \textit{\revision{Forced Action}} & \textbf{0.89} & 0.85 & 0.78 & 0.79 & 0.83 & \textbf{0.88} & \textbf{0.84} & \textbf{0.84} & 0.82 \\
& \cellcolor{aliceblue}\textit{\revision{Hidden Costs}} & \cellcolor{aliceblue}\textbf{0.60} & \cellcolor{aliceblue}- & \cellcolor{aliceblue}0.05 & \cellcolor{aliceblue}\textbf{0.60} & \cellcolor{aliceblue}- & \cellcolor{aliceblue}0.20 & \cellcolor{aliceblue}\textbf{0.60} & \cellcolor{aliceblue}- & \cellcolor{aliceblue}0.08 \\
& \textit{\revision{\begin{tabular}[c]{@{}l@{}}Hidden \\ Subscription\end{tabular}}} & \textbf{0.90} & 0.78 & 0.56 & \textbf{0.95} & 0.72 & 0.77 & \textbf{0.92} & 0.75 & 0.64 \\
& \cellcolor{aliceblue}\textit{\revision{Nudge}} & \cellcolor{aliceblue}\textbf{0.74} & \cellcolor{aliceblue}0.57 & \cellcolor{aliceblue}0.17 & \cellcolor{aliceblue}\textbf{0.80} & \cellcolor{aliceblue}0.37 & \cellcolor{aliceblue}0.68 & \cellcolor{aliceblue}\textbf{0.77} & \cellcolor{aliceblue}0.45 & \cellcolor{aliceblue}0.28 \\
& \textit{\revision{Pre-Selection}} & \textbf{0.67} & 0.65 & 0.33 & \textbf{1.00} & \textbf{1.00} & \textbf{1.00} & \textbf{0.80} & 0.79 & 0.50 \\
& \cellcolor{aliceblue}\textit{\revision{Trick Wording}} & \cellcolor{aliceblue}\textbf{0.86} & \cellcolor{aliceblue}0.61 & \cellcolor{aliceblue}0.34 & \cellcolor{aliceblue}\textbf{0.80} & \cellcolor{aliceblue}0.61 & \cellcolor{aliceblue}0.47 & \cellcolor{aliceblue}\textbf{0.83} & \cellcolor{aliceblue}0.61 & \cellcolor{aliceblue}0.39 \\
& \textit{\revision{\begin{tabular}[c]{@{}l@{}}Visual \\ Interference\end{tabular}}} & \textbf{0.50} & 0.36 & 0.06 & 0.83 & 0.80 & \textbf{1.00} & \textbf{0.62} & 0.50 & 0.11 \\
\cmidrule(lr){1-11}
& Deceptive & \textbf{0.89} & 0.84 & 0.65 & \textbf{0.95} & 0.80 & \textbf{0.95} & \textbf{0.92} & 0.82 & 0.77 \\
& Not Deceptive & \textbf{1.00} & 0.98 & 0.99 & \textbf{0.99} & \textbf{0.99} & 0.95 & \textbf{0.99} & 0.98 & 0.97 \\
\bottomrule
\end{tabular}
}
\label{tab:langeval_no_t5base}
\end{table}

These results highlight the trade-offs between the different language models. As expected, Gemini exhibits the highest performance, reaching near perfect precision and recall on most deceptive patterns. It struggles in two pattern subtypes:  hidden-costs and visual interference. In second place is the Qwen model, which struggles in more subtypes. The distilled \texttt{T5} model exhibits generally acceptable performance, but struggles for pattern categories: interface-interference and obstruction.

\section{End-to-End Evaluation}
\label{sec:evaluation}

We perform an end-to-end evaluation of \name on a real-world dataset, consisting of 1.1K websites, manually annotated by 2 authors (described in~\Cref{sub:eval:dataset-creation}). We perform our evaluation with all three LLMs in the \textit{LanguageModule} (see~\Cref{sub:possible-solutions}). For this End-to-End Evaluation, our objective is to measure the accuracy of our entire framework (including the vision and language models) in detecting deceptive patterns on websites.

\subsection{Dataset}
\label{sub:eval:dataset-creation}

To create the dataset for the end-to-end evaluation, we crawl the websites in the deceptive pattern dataset from Mathur et al.~\cite{mathur2019dark}.  We use this dataset since it is the latest and most comprehensive dataset of websites with deceptive patterns. We filter out the non-English and offline websites from this dataset, leaving us with 555 websites out of the 1400 sites mentioned. Next, we crawled the top 1000 websites from the Tranco~\cite{pochat2018tranco} list\footnote{\url{https://tranco-list.eu/list/QGJ74/1000000}}, utilizing the same methodology in \Cref{subsub:d3-dataset} to filter out non-English and NFSW websites, resulting in 597 websites. \revision{Additionally, for the Tranco websites, we did not scrape just the landing pages of websites. Rather, we queried “site:domain” on DuckDuckGo and programmatically counted the number of interactable elements for each of the top ten resulting pages. The page with the highest count was selected for analysis.}
In total, we have 1152 websites in our end-to-end evaluation dataset.

Next, we manually labeled the screenshots of these websites to identify and localize the deceptive patterns, associating each screenshot with an \textmap. Specifically, two authors annotated the same randomly selected 100 websites with high inter-annotator agreement ($\kappa = 0.89$)~\cite{landis1977measurement}. Then, each researcher separately labeled the rest of the screenshots. We refer to this dataset as the golden Deceptive Design Dataset (\textit{D3 dataset}). We show the distribution of the samples in \textit{D3 Dataset} in \Cref{tab:d3}.

\begin{table}[h]
\caption{Distribution of Samples in \textit{D3 Dataset}.}
\centering
\resizebox{0.6\columnwidth}{!}{
\begin{tabular}{l l c}
\toprule
\textbf{Category} & \textbf{Subtype} & \textbf{\# Samples} \\
\midrule
\textbf{\revision{Non Deceptive}}  & \textit{\revision{Not Applicable}}  & 22618 \\
\cmidrule(lr){1-3}
\textbf{\revision{Forced Action}}  & \textit{\revision{Forced Action}}  & 414 \\
\cmidrule(lr){1-3}
\multirow{3}{*}{\centering \textbf{\revision{Interface Interference}}} & \textit{\revision{Nudge}} & 143 \\
& \textit{\revision{Fake Scarcity / Fake Urgency}} & 211 \\
& \textit{\revision{Confirmshaming}} & 42 \\
\cmidrule(lr){1-3}
\multirow{2}{*}{\centering \textbf{\revision{Obstruction}}} & \textit{\revision{Visual Interference}} & 48 \\
& \textit{\revision{Pre-Selection}} & 18 \\
\cmidrule(lr){1-3}
\multirow{4}{*}{\centering \textbf{\revision{Sneaking}}} & \textit{\revision{Trick Wording}} & 190 \\
& \textit{\revision{Hidden Costs}} & 28 \\
& \textit{\revision{Hidden Subscription}} & 379 \\
& \textit{\revision{Disguised Ads}} & 306 \\
\bottomrule
\end{tabular}
}
\label{tab:d3}
\end{table}

\begin{table*}[t!]
  \caption{\revision{Performance of \name (with three underlying language models: Gemini, Qwen, and T5) and \dpguard{DPGuard}~\cite{50Shades} on the \textit{D3 Dataset}. The table has three sections, each showing performance at the category, subtype, and binary levels.}}
  \label{tab:model_performance_subtype}
  \centering
  \resizebox{\textwidth}{!}{%
    \begin{threeparttable}
      \begin{tabular}{@{}llcccccccccccc@{}}
        \toprule
        & \multirow{2}{*}{\textbf{Pattern Type}}
        & \multicolumn{4}{c}{\textbf{Precision}}
        & \multicolumn{4}{c}{\textbf{Recall}}
        & \multicolumn{4}{c}{\textbf{F1‑Score}} \\
        \cmidrule(lr){3-6} \cmidrule(lr){7-10} \cmidrule(lr){11-14}
        & 
        & \textbf{Gemini} & \textbf{Qwen}  & \textbf{T5}  & \textbf{\dpguard{DPGuard}}\revision{~\cite{50Shades}}
        & \textbf{Gemini} & \textbf{Qwen}  & \textbf{T5}  & \textbf{\dpguard{DPGuard}}\revision{~\cite{50Shades}}
        & \textbf{Gemini} & \textbf{Qwen}  & \textbf{T5}  & \textbf{\dpguard{DPGuard}}\revision{~\cite{50Shades}} \\
        \midrule
        \multirow{5}{*}{\cellcolor{white}\rotatebox{90}{\textbf{Category}}}
        & \cellcolor{aliceblue}\textbf{\revision{Non Deceptive}}
          & \cellcolor{aliceblue}\textbf{1.00} & \cellcolor{aliceblue}0.98 & \cellcolor{aliceblue}\textbf{1.00} & \cellcolor{aliceblue}–
          & \cellcolor{aliceblue}\textbf{0.99} & \cellcolor{aliceblue}\textbf{0.99} & \cellcolor{aliceblue}0.96 & \cellcolor{aliceblue}–
          & \cellcolor{aliceblue}\textbf{0.99} & \cellcolor{aliceblue}\textbf{0.99} & \cellcolor{aliceblue}0.98 & \cellcolor{aliceblue}– \\
        & \textbf{\revision{Forced Action}}
          & \textbf{0.97} & 0.89 & 0.82 & –
          & \textbf{0.94} & 0.85 & 0.83 & –
          & \textbf{0.95} & 0.87 & 0.82 & – \\
        & \cellcolor{aliceblue}\textbf{\revision{Interface Interference}}
          & \cellcolor{aliceblue}\textbf{0.89} & \cellcolor{aliceblue}0.76 & \cellcolor{aliceblue}0.55 & \cellcolor{aliceblue}–
          & \cellcolor{aliceblue}\textbf{0.95} & \cellcolor{aliceblue}0.64 & \cellcolor{aliceblue}0.78 & \cellcolor{aliceblue}–
          & \cellcolor{aliceblue}\textbf{0.92} & \cellcolor{aliceblue}0.69 & \cellcolor{aliceblue}0.64 & \cellcolor{aliceblue}– \\
        & \textbf{\revision{Obstruction}}
          & \textbf{0.70} & 0.57 & 0.13 & –
          & \textbf{0.97} & 0.71 & 0.93 & –
          & \textbf{0.81} & 0.63 & 0.23 & – \\
        & \cellcolor{aliceblue}\textbf{\revision{Sneaking}}
          & \cellcolor{aliceblue}\textbf{0.87} & \cellcolor{aliceblue}0.79 & \cellcolor{aliceblue}0.52 & \cellcolor{aliceblue}–
          & \cellcolor{aliceblue}\textbf{0.94} & \cellcolor{aliceblue}0.73 & \cellcolor{aliceblue}0.85 & \cellcolor{aliceblue}–
          & \cellcolor{aliceblue}\textbf{0.90} & \cellcolor{aliceblue}0.76 & \cellcolor{aliceblue}0.64 & \cellcolor{aliceblue}– \\
        \cmidrule(lr){1-14}
        \multirow{11}{*}{\cellcolor{white}\rotatebox{90}{\textbf{Subtype}}}
        & \textit{\revision{Not Applicable}}
          & \textbf{1.00} & 0.98 & \textbf{1.00} & 0.78
          & \textbf{0.99} & \textbf{0.99} & 0.91 & 0.74
          & \textbf{0.99} & \textbf{0.99} & 0.95 & 0.76 \\
        & \cellcolor{aliceblue}\textit{\revision{Confirmshaming}}
          & \cellcolor{aliceblue}\textbf{0.93} & \cellcolor{aliceblue}0.75 & \cellcolor{aliceblue}0.59 & \cellcolor{aliceblue}0.05
          & \cellcolor{aliceblue}\textbf{1.00} & \cellcolor{aliceblue}0.69 & \cellcolor{aliceblue}0.85 & \cellcolor{aliceblue}0.32
          & \cellcolor{aliceblue}\textbf{0.97} & \cellcolor{aliceblue}0.72 & \cellcolor{aliceblue}0.70 & \cellcolor{aliceblue}0.09 \\
        & \textit{\revision{Disguised Ads}}
          & \textbf{0.74} & 0.62 & 0.34 & 0.49
          & \textbf{0.95} & 0.61 & 0.81 & 0.62
          & \textbf{0.83} & 0.61 & 0.48 & 0.55 \\
        & \cellcolor{aliceblue}\textit{\revision{Fake Scarcity / Fake Urgency}}
          & \cellcolor{aliceblue}\textbf{0.94} & \cellcolor{aliceblue}0.87 & \cellcolor{aliceblue}0.68 & \cellcolor{aliceblue}–
          & \cellcolor{aliceblue}\textbf{0.96} & \cellcolor{aliceblue}0.71 & \cellcolor{aliceblue}0.89 & \cellcolor{aliceblue}–
          & \cellcolor{aliceblue}\textbf{0.95} & \cellcolor{aliceblue}0.78 & \cellcolor{aliceblue}0.77 & \cellcolor{aliceblue}– \\
        & \textit{\revision{Forced Action}}
          & \textbf{0.97} & 0.89 & 0.72 & 0.40
          & \textbf{0.94} & 0.85 & 0.84 & 0.51
          & \textbf{0.95} & 0.87 & 0.77 & 0.45 \\
        & \cellcolor{aliceblue}\textit{\revision{Hidden Costs}}
          & \cellcolor{aliceblue}\textbf{0.77} & \cellcolor{aliceblue}0.31 & \cellcolor{aliceblue}0.14 & \cellcolor{aliceblue}–
          & \cellcolor{aliceblue}\textbf{0.91} & \cellcolor{aliceblue}0.28 & \cellcolor{aliceblue}0.38 & \cellcolor{aliceblue}–
          & \cellcolor{aliceblue}\textbf{0.83} & \cellcolor{aliceblue}0.29 & \cellcolor{aliceblue}0.21 & \cellcolor{aliceblue}– \\
        & \textit{\revision{Hidden Subscription}}
          & \textbf{0.93} & 0.84 & 0.51 & –
          & \textbf{0.96} & 0.72 & 0.83 & –
          & \textbf{0.95} & 0.78 & 0.63 & – \\
        & \cellcolor{aliceblue}\textit{\revision{Nudge}}
          & \cellcolor{aliceblue}\textbf{0.79} & \cellcolor{aliceblue}0.52 & \cellcolor{aliceblue}0.10 & \cellcolor{aliceblue}0.23
          & \cellcolor{aliceblue}\textbf{0.89} & \cellcolor{aliceblue}0.43 & \cellcolor{aliceblue}0.48 & \cellcolor{aliceblue}0.58
          & \cellcolor{aliceblue}\textbf{0.83} & \cellcolor{aliceblue}0.47 & \cellcolor{aliceblue}0.17 & \cellcolor{aliceblue}0.33 \\
        & \textit{\revision{Pre-Selection}}
          & \textbf{0.64} & 0.64 & 0.25 & 0.02
          & 0.94 & 0.94 & \textbf{1.00} & 0.14
          & \textbf{0.76} & \textbf{0.76} & 0.40 & 0.03 \\
        & \cellcolor{aliceblue}\textit{\revision{Trick Wording}}
          & \cellcolor{aliceblue}\textbf{0.85} & \cellcolor{aliceblue}0.74 & \cellcolor{aliceblue}0.57 & \cellcolor{aliceblue}0.00
          & \cellcolor{aliceblue}\textbf{0.82} & \cellcolor{aliceblue}0.69 & \cellcolor{aliceblue}0.67 & \cellcolor{aliceblue}0.00
          & \cellcolor{aliceblue}\textbf{0.83} & \cellcolor{aliceblue}0.71 & \cellcolor{aliceblue}0.62 & \cellcolor{aliceblue}0.00 \\
        & \textit{\revision{Visual Interference}}
          & \textbf{0.73} & 0.54 & 0.03 & 0.09
          & \textbf{0.98} & 0.62 & 0.93 & 0.22
          & \textbf{0.84} & 0.58 & 0.05 & 0.13 \\
        \cmidrule(lr){1-14}
        & Deceptive
          & \textbf{0.90} & 0.88 & 0.72 & –
          & \textbf{0.97} & 0.81 & 0.95 & –
          & \textbf{0.93} & 0.84 & 0.82 & – \\
        & Not Deceptive
          & \textbf{1.00} & 0.98 & 0.97 & –
          & \textbf{0.99} & \textbf{0.99} & 0.97 & –
          & \textbf{0.99} & \textbf{0.99} & 0.98 & – \\
        \bottomrule
      \end{tabular}
      \begin{tablenotes}
        \item[]–: the DP classification is not supported by the model.
      \end{tablenotes}
    \end{threeparttable}%
  }
\end{table*}

\subsection{Findings} 
We analyze the websites in the \textit{D3 Dataset} using \name and compare the predicted deceptive patterns to the ground truth annotations.

The results from the evaluation are shown in~\Cref{tab:model_performance_subtype}. We observe that the performance of the \name framework is mainly dependent on the type of language model used. We note here that \texttt{Gemini} is the teacher model in our framework, and \texttt{Qwen} and \texttt{T5} are the student models, as described in \Cref{sub:possible-solutions}. As such, we see that Gemini has the best performance, followed by \texttt{Qwen} and \texttt{T5}. 

We also evaluate Shi et al.'s~\cite{50Shades} \textit{DPGuard} framework. To evaluate their framework, we first create a mapping between their taxonomy and the filtered taxonomy (see~\Cref{app:dpguard-mapping}). For our evaluation, we only consider the categories mappable to the filtered taxonomy. Additionally, as the \textit{DPGuard} framework does not provide localization capabilities, our evaluation is solely focused on \textit{DPGuard}'s ability to identify deceptive patterns present within the webpage in our \textit{D3 Dataset}. Our evaluation shows that \textit{DPGuard} is unable to identify deceptive patterns with high accuracy, especially struggling to classify \textit{`hidden-subscription'} and \textit{`trick-wording'}. These findings are consistent with the evaluation performed by Shi et al.~\cite{50Shades}.







\subsubsection{Error Analysis} In the end-to-end evaluation, we observe that, overall, \texttt{Gemini} performs extremely well in identifying deceptive patterns. However, for certain subtypes, it has comparatively lower performance. For instance, \texttt{Gemini} has reduced performance in identifying \textit{`pre-selection'}. Similarly, for \textit{`disguised-ads'}, it has a slightly higher false positive rate. Further analysis shows that these false positives are mainly due to product placement: typically benign content resembling deceptive advertising. For instance, a website showcasing template-based shopping apps may include screenshots of user-created apps, some of which contain promotional text. Although harmless, this text is often misclassified as \textit{`disguised-ads'} due to its advertising-like appearance. Furthermore, through empirical analysis, we have observed \texttt{Gemini} interchangeably using \textit{`nudge'} and \textit{`forced-action'}, especially when classifying cookie notices. 

For the smaller models, \texttt{Qwen} and \texttt{T5}, we observe that the \name sometimes fails to identify deceptive pattern categories/subtypes correctly. Investigating the error cases further, we find instances where \name misclassified the category or sub-category of the deceptive pattern (while correctly identifying that the pattern is deceptive). For example, there are instances where \name incorrectly identifies \textit{forced-action} as \textit{nudge}. We also find that on Wikipedia, \name incorrectly tags a non-deceptive pattern as deceptive. The classification text contains money or cost-related text, causing the model to classify the text as \textit{trick-wording} incorrectly. 
\smallskip

\subsubsection{Impact of Errors} We note that the impact of the misclassifications due to category or sub-category mismatch is minimal on the users as the users will still be notified of a deceptive pattern. For false positives, the user gets notified for a pattern where none exists. Upon further inspection, users can safely ignore the notification, causing minimal distraction. For false negatives - the user impact can be severe. In such cases, users might get into a sense of false security and get deceived by the deceptive pattern because of \name's error. We emphasize that high recall of \name ensures that such cases will be minimal.

\section{Applications}
\label{sec:applications}

We instantiate the \name framework across three potential downstream tasks, each designed to serve a stakeholder in the web ecosystem: users, developers, and regulators and researchers.

\subsection{Browser Extension for Web Users}
\label{subsec:browser_extension}
\begin{figure}[h]
    \centering
    \subfloat[Screenshot of a website.]{
        \includegraphics[width=0.48\columnwidth]{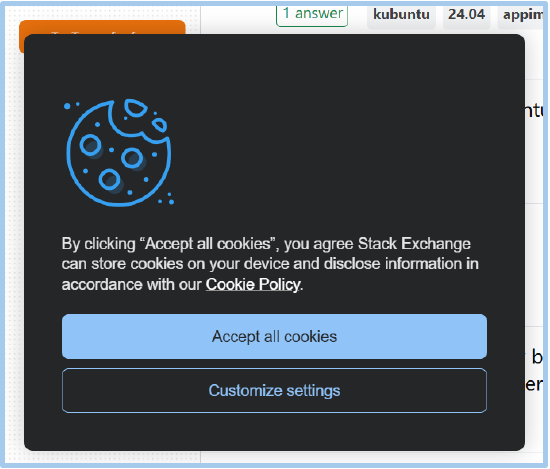}
        \label{fig:figure1}
    }
    \subfloat[Screenshot of \name running on browser and detecting deceptive patterns.]{
        \includegraphics[width=0.48\columnwidth]{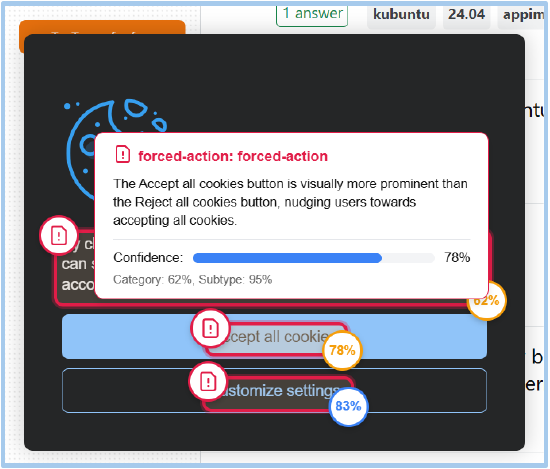}
        \label{fig:figure2}
    }
    \caption{Screenshot of website (left) and \name running (right).}
    \label{fig:combined}
\end{figure}
Our first instantiation is a browser extension that directly helps web users, the primary target of deceptive patterns. The extension takes a screenshot of the user's active page and analyzes it using the \name framework. The active page may contain sensitive information of the user. To mitigate data privacy concerns, the extension performs the analysis locally using a distilled version of \texttt{Flan-T5}~\Cref{subsec:distil-t5}. Once processed, the extension highlights deceptive patterns to the users as shown in \Cref{fig:combined}. Specifically, it shows bounding boxes around the UI elements where deceptive patterns are found, and informs the users as they hover over elements.

\paragraph{Extension Architecture}
The browser extension utilizes a hybrid architecture consisting of lightweight browser components with a locally running Flask server to run the \name framework, similar to the \textit{Zotero} extension~\cite{zotero}.
On the browser side, we use a popup script to allow users to activate the extension. Once active, the extension takes a screenshot and sends it to the local Flask server for analysis. The Flask server hosts the \name framework, using \texttt{Flan-T5} as the language model. After analysis, the server returns the classifications to the extension, which are then rendered by the content script of the extension on the user's screen. This rendering is shown in~\Cref{fig:combined}.

\paragraph{System Level Performance} For the browser extension to be practical, it must perform its analysis in real time. We, therefore, conducted latency tests on its core \texttt{Flan-T5} (Language) and \texttt{YOLOv10 Ensemble} (Vision) models using three different machine configurations representing various hardware capabilities (modern high-end with GPU, older mid-range, ARM-based) and comparing CPU versus GPU performance. Our results (detailed in~\Cref{app:latency}) show that while performance on older hardware using only the CPU takes roughly \textbf{1.5 to 3 seconds} per module (e.g., \textbf{1.95 seconds} for \texttt{YOLOv10 Ensemble} on a CPU-only i5 laptop), newer devices with GPUs achieve near real-time performance. Specifically, on the 2023 laptop with a GPU, \texttt{T5} inference (with 800 tokens as input) took less than \textbf{0.5 seconds}, and \texttt{YOLOv10 Ensemble} performed its classifications in less than \textbf{0.3 seconds}. Our experiments show the feasibility of implementing \name as a browser extension, as modern hardware with GPU support enables a responsive experience without major performance delays~\cite{huggingface_transformersjs_webgpu,w3c_webgpu_2024}.

\subsection{Lighthouse Reports for Web Developers}
\label{subsec:lighthouse}

\begin{figure}[ht]
    \centering
  \includegraphics[width=0.8\columnwidth]{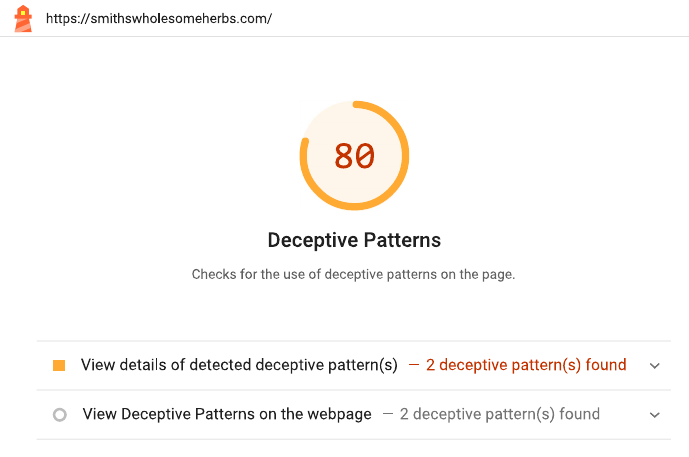}
    \caption{Custom Audit on a Lighthouse Report.}
    \label{fig:lighthouse-report}
\end{figure}

Research shows that while developers are often unaware of deceptive patterns on their websites and their impacts on users, they are open to addressing these patterns if properly informed~\cite{stover_how_2023}. To that end, we integrate \name into Lighthouse~\cite{lighthouse_chrome,lighthouse_overview}, a popular tool developers use to receive automated audits on website quality. Google bundles this tool with ChromeDev Tools, and major platforms such as Shopify, Wix, and Squarespace\footnote{\url{https://www.shopify.com/}, \url{https://www.wix.com/}, \url{https://www.squarespace.com/}} integrate it into their workflows. We created a custom Lighthouse audit using \name to fit directly into a web developer's workflow. This audit uses an existing \texttt{Lighthouse Gatherer}~\cite{lighthouse_gatherer} to capture screenshots of the website, processes it using \name, and finally incorporates the findings into a Lighthouse report, as shown in \Cref{fig:lighthouse-report}.

We package Lighthouse with our custom audit in a docker container for better usability. The docker container maintains all functionality of Lighthouse while adding our custom audit. The container simply takes an URL as input to run the entire audit and give the developer a report. The report provides the developer with \textit{DeceptivePattern Score} based on the number of deceptive patterns ($n$) found on the page using the following scoring scheme: 
\[
S(n) =
\begin{cases}
100, & \text{if } n = 0 \\
89, & \text{if } n = 1 \\
\max(100 - 10n, 0), & \text{if } n > 1
\end{cases}
\]

The scoring scheme scales inversely with the number of deceptive patterns. It assigns a score of $0.89$ for a single deceptive pattern to ensure that the audit shows a failure condition. An example of the Lighthouse Report as a developer would see is shown in~\Cref{fig:lighthouse-report}.

\subsection{Enabling Web-Scale Analysis}
\label{subsec:measurements}
Our last instantiation of the \name framework is a tool designed for scalable website analysis. This tool serves researchers and regulators, providing them with the necessary automated capabilities to investigate the broader landscape of online deceptive practices. It takes input from a list of URLs. Then, it crawls each URL, takes a screenshot, extracts the \textmap from each screenshot, and passes the \textmap to the Language Module. We use the Gemini model with batch API in this tool to generate accurate analysis at a low cost.

\paragraph{Measurement on Shopify and Tranco Websites}
We use this tool to analyze 11,118 websites, consisting of 6,626 diverse, popular domains selected from the Tranco list~\cite{pochat2018tranco}, and 4,492 e-commerce sites from the \textit{Shopify Partners Directory}\footnote{\url{https://www.shopify.com/partners/directory}}. We detail our website selection process earlier in ~\Cref{subsub:d3-dataset}.

\begin{figure}[ht]
    \centering
    \begin{subfigure}{0.9\columnwidth}
        \centering
        \includegraphics[width=0.9\columnwidth]{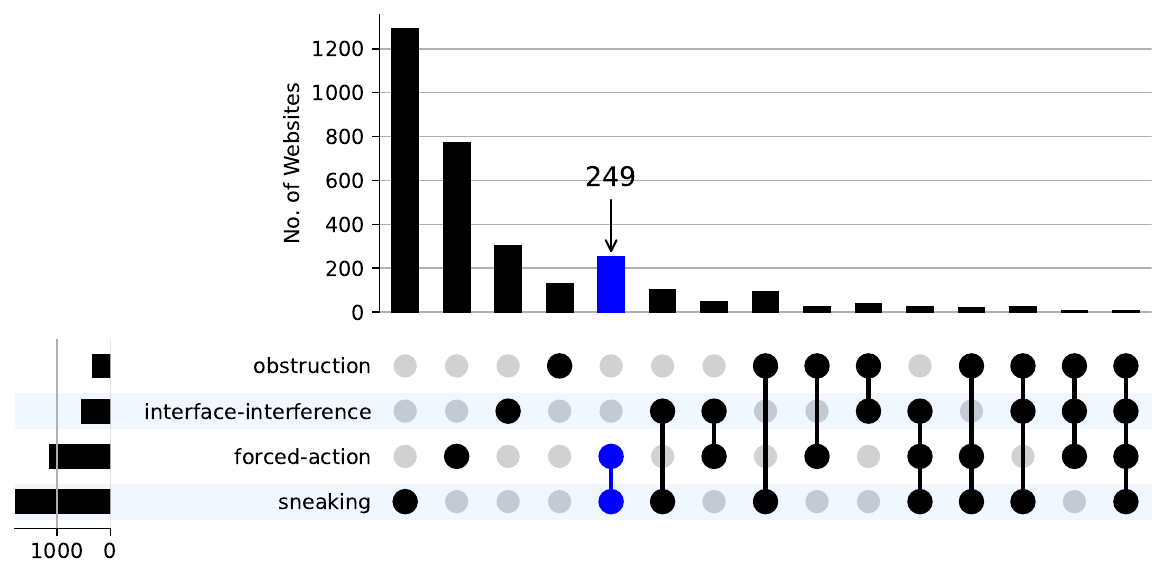}
        \caption{Distribution of Deceptive Patterns identified by \name on Tranco websites}
        \label{fig:upset-tranco}
    \end{subfigure}
    
    \vspace{1em}  

    \begin{subfigure}{0.9\columnwidth}
        \centering
        \includegraphics[width=0.9\columnwidth]{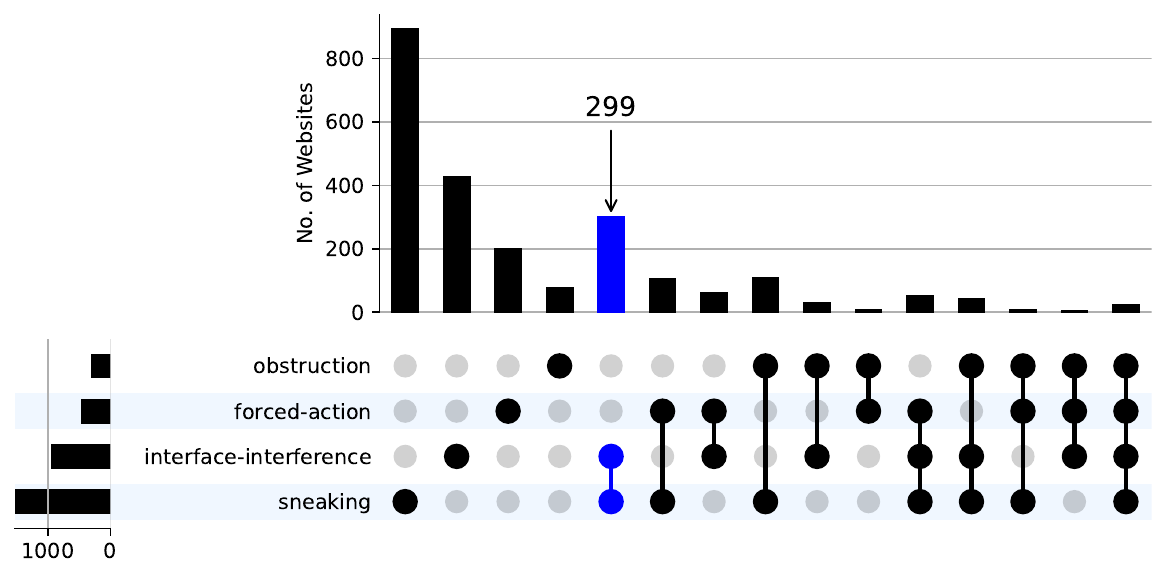}
        \caption{Distribution of Deceptive Patterns identified by \name on Shopify Websites}
        \label{fig:upset-shopify}
    \end{subfigure}
    
    \caption{Distribution of deceptive patterns across various domains such as (a) Most visited Tranco websites and (b) Shopify based e-commerce websites}
    \label{fig:upset}
\end{figure}

The complete distribution of the identified deceptive patterns across the two domains is in~\Cref{fig:upset}. We observe that \textit{sneaking} is the most prevalent deceptive pattern across both domains. On Shopify websites, the second prevalent is \textit{interface-interference}. For example, e-commerce websites tend to use fake scarcity or urgency (e.g., ``Limited time offer: get 20\% off for next 5 min''), which is classified as \textit{fake-scarcity-fake-urgency}. On Tranco websites, \textit{forced-action} is the next most prevalent pattern after \textit{sneaking}. Our analysis also reveals that many websites employ multiple distinct categories of deceptive patterns simultaneously. For example, \url{bedbathandbeyond.com} (\Cref{app:bbb}) consists of \textit{`forced-action'} (by providing no clear option to reject cookies) alongside `\textit{obstruction}' (by presenting mailing list terms and conditions in an excessively small font).


\begin{figure}[H]
    \centering
    \includegraphics[width=0.45\textwidth]{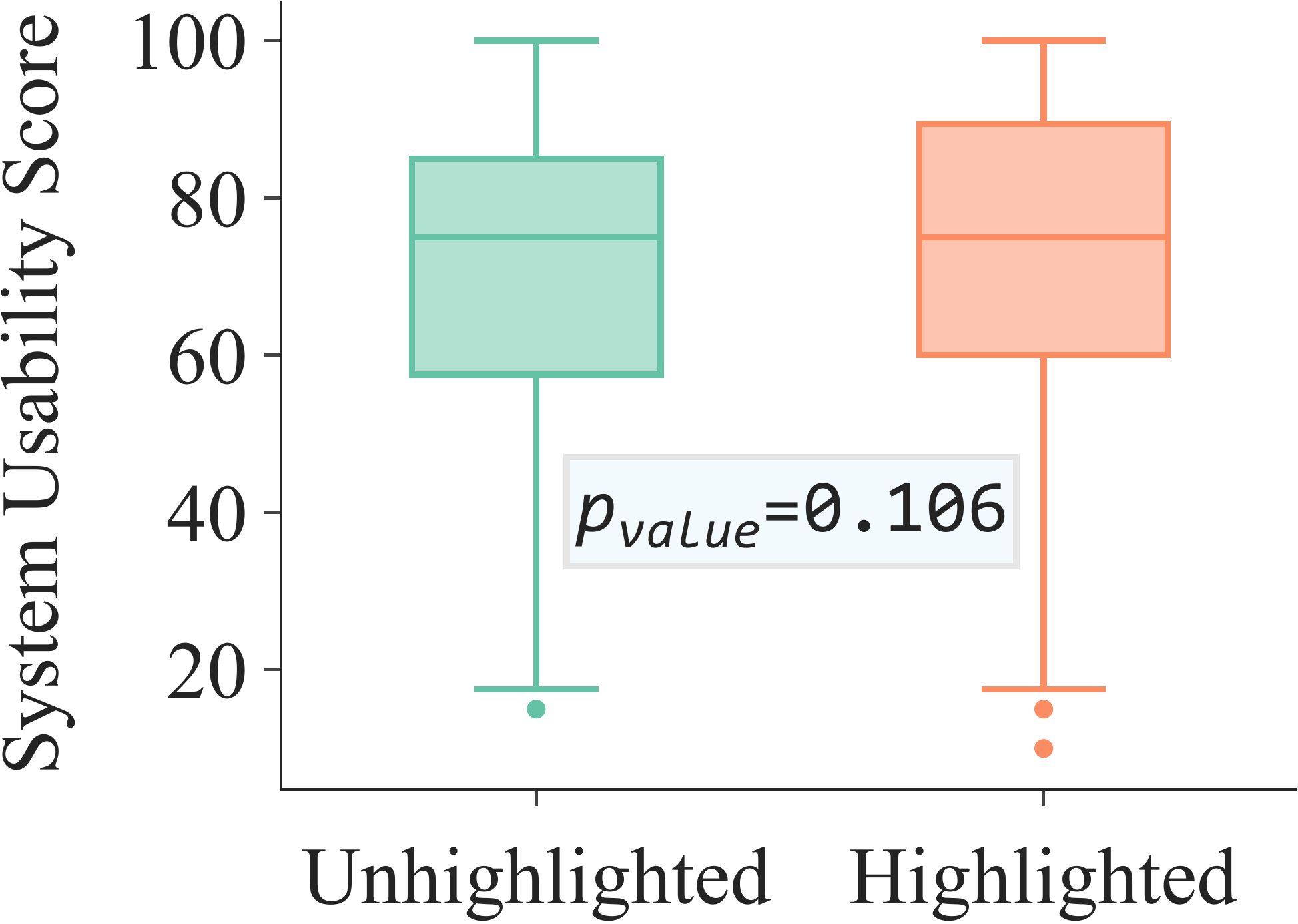}
    \caption{The results from the usability study show how highlighting affected the website's user experience. We find that usability is not affected by the highlights $(p=0.106)$.}
    \label{fig:user_sus}
\end{figure}

\begin{figure}[b]
    \centering
    \includegraphics[width=0.15\textwidth]{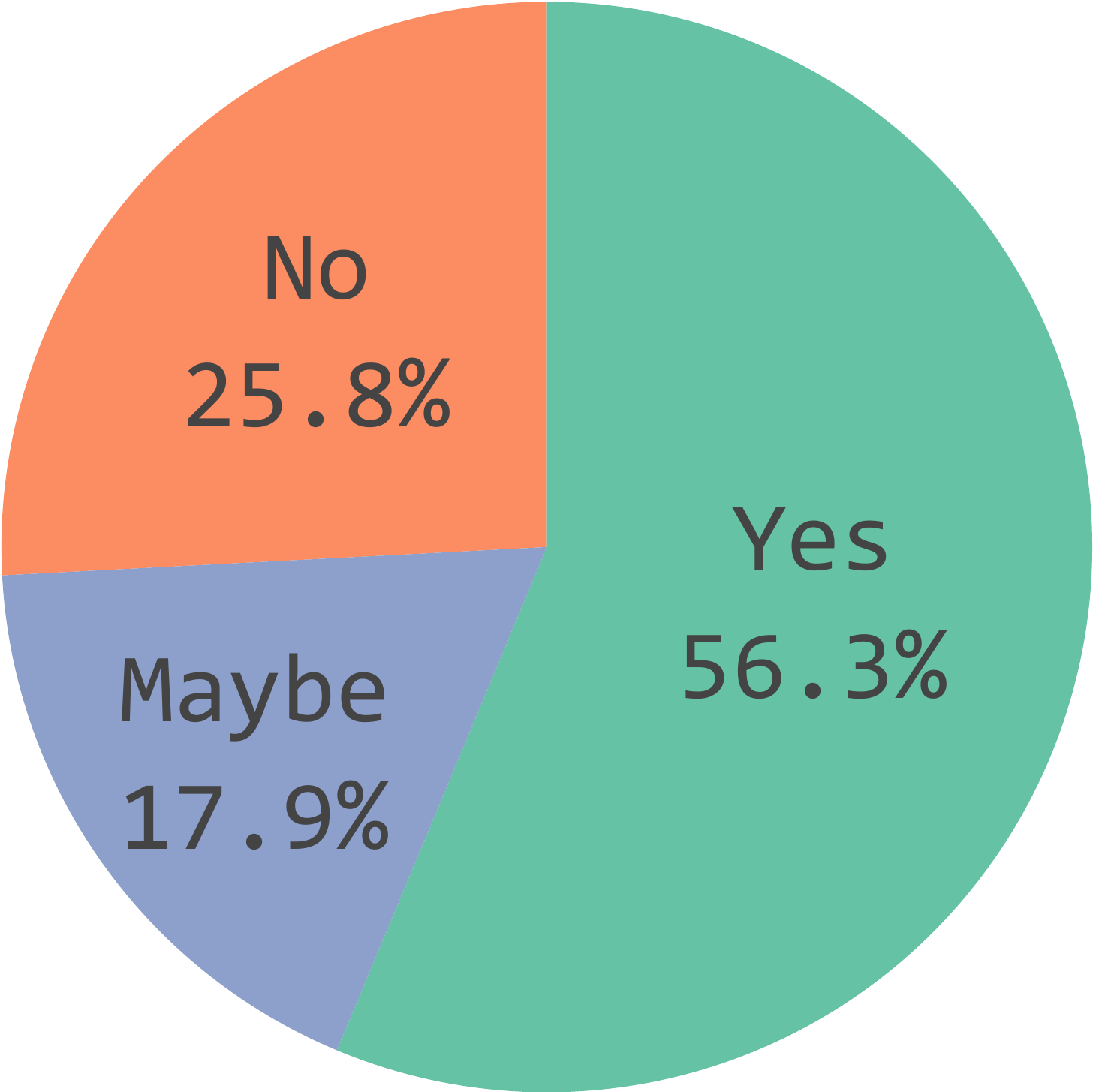}
    \hfill
    \includegraphics[width=0.15\textwidth]{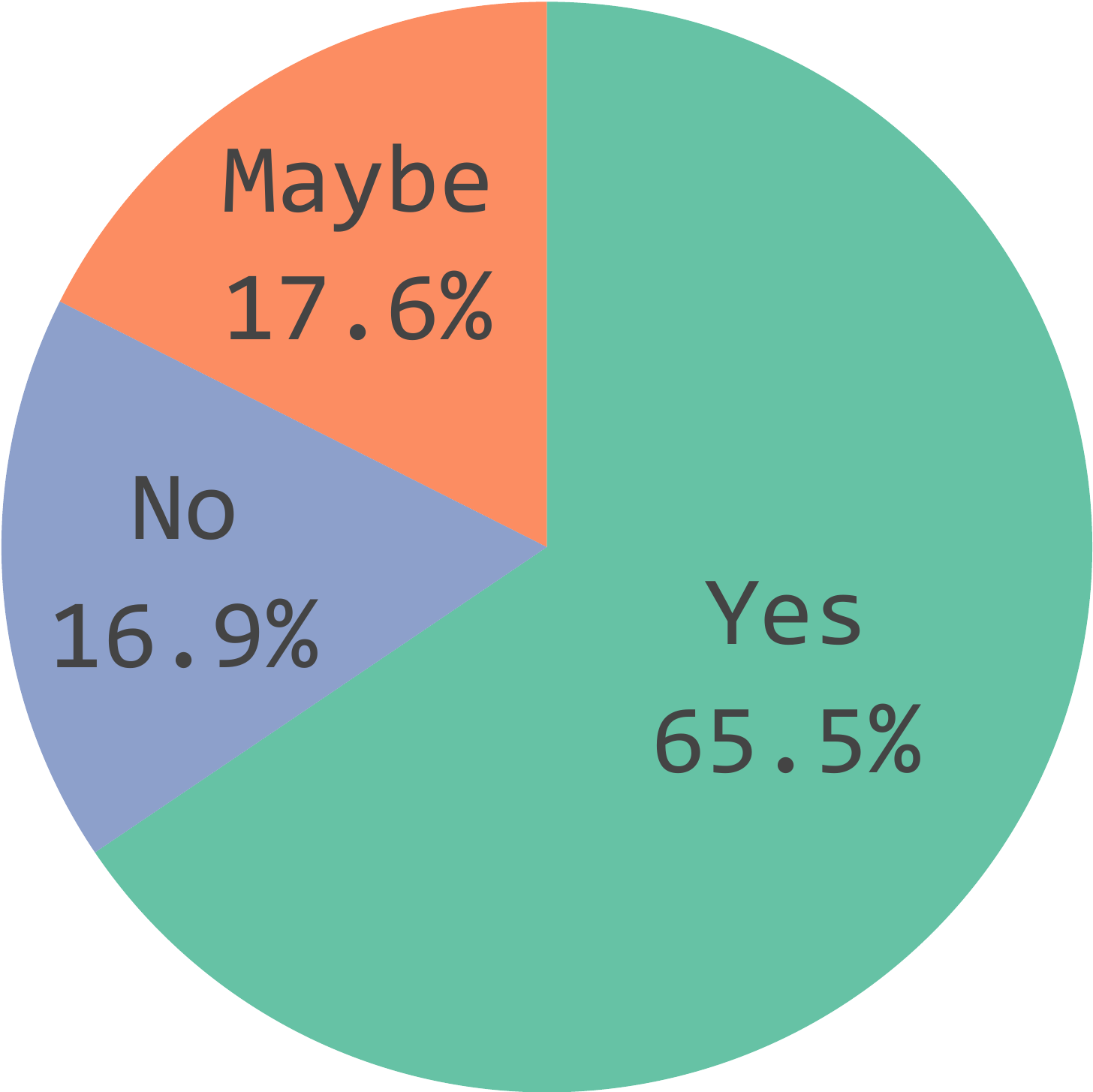}
    \hfill
    \includegraphics[width=0.15\textwidth]{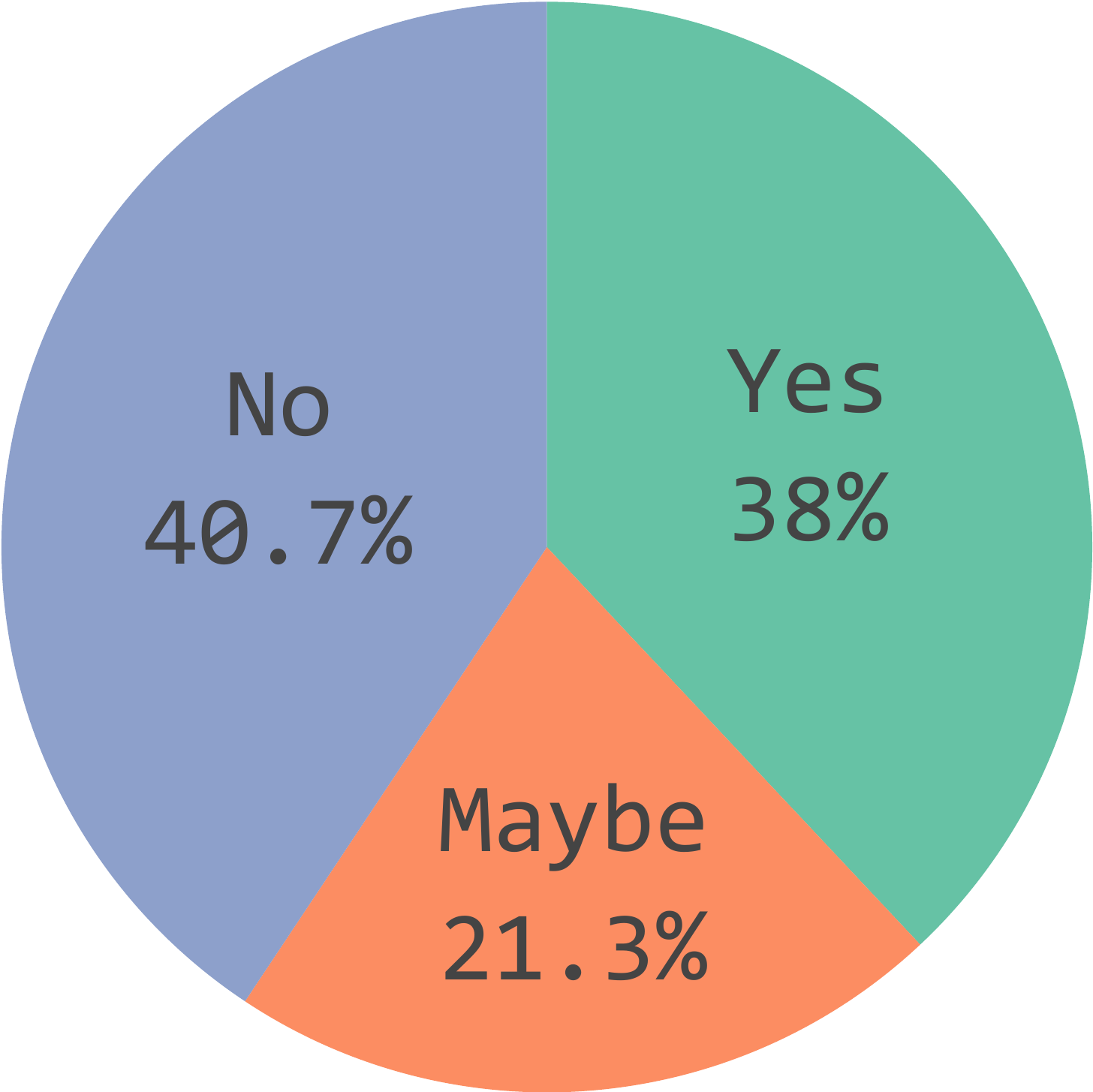}
    \caption{The distribution of participant responses to qualitative Q1-3 asked after the user study.}
    \label{fig:user_resp}
\end{figure}

\revision{\section{User Studies}

We conducted two exploratory user studies to evaluate the usability of two \name applications: browser extension and lighthouse report. In our first study, we evaluate the usability of a website with highlighted deceptive patterns. Next, we reached out to Shopify developers from the Shopify developer list (see \Cref{subsec:measurements}), with the Lighthouse report of their website.

\subsection{Website Usability}

We perform an exploratory user-based evaluation to explore how highlighting deceptive patterns affects the usability of a website. We recruited 151 U.S.-based participants from Prolific, compensating them \$2 for a task with a median completion time of 7 minutes. We did not collect demographic data and asked Prolific to distribute the survey evenly across the demographics. The IRB at our institution determined that the proposed activity is not research involving human subjects as defined by DHHS and FDA regulations.

\subsubsection{Study Design}

We conducted a within-subject study to assess the usability impact of highlighting deceptive patterns with \name. We informed the participants that the objective of the survey was to test the usability of a web interface. Next, participants were asked to visit two custom-made websites, one with and one without highlighting, and fill out a System Usability Scale questionnaire~\cite{brooke1996sus} after each. In both these websites, the participants were asked to perform one of the four tasks: sign up, download, do shopping, or read a news article. Once the participants visit a website and hover over the highlighted text, a banner cautioning them about the deceptive pattern is shown. Note that the participants interacted with highlighted websites directly without installing a browser extension. 

These custom-made websites were created by the authors based on various real websites, available at \textit{UXP\textsuperscript{2} Dark Patterns}\footnote{\url{https://darkpatterns.uxp2.com/}}, caught using deceptive patterns. An example of such a website is shown in \Cref{app:mock-sites}. After visiting each website, the participant is asked to complete a System Usability Scale questionnaire~\cite{brooke1996sus}. Additionally, we asked the participants to fill out a post-study questionnaire consisting of three questions: 
\begin{enumerate}[nosep]
    \item [Q1.] \textit{Did you find the hints about the deceptive patterns useful? }
    \item [Q2.] \textit{Did you feel that the highlighted box encouraged you to notice the deceptive patterns?}
    \item [Q3.] \textit{Did you feel that the highlighted box encouraged you to change your choice?}
\end{enumerate}


\subsubsection{Findings}

We evaluated the website's usability with and without highlights using the SUS metric, , as shown in \Cref{fig:user_sus}. In this case, we consider the non-hypothesis to be that the website's usability is unaffected after highlighting. Based on the Wilcoxon signed-rank statistical test, the $p$-value for comparing two groups is $0.106$, indicating no statistically significant difference between the SUS scores for websites. Therefore, the null hypothesis stands that usability is unaffected. We do note, however, that the mean SUS score for the highlighted website was 2 points higher.

Furthermore, based on participant feedback to the three questions mentioned above (\Cref{fig:user_resp}), we find that most found the highlight box (65.5\%) and its corresponding hint (56.3\%) helpful for recognizing deceptive patterns. Additionally, 38\% of the participants reported they would change their behavior after being shown the highlighted patterns.


\subsubsection{Future Studies} Reflecting on our preliminary user study results suggests that the design of highlighting deceptive patterns on webpages does not affect website usability, and can help users recognize deceptive patterns. Follow-up studies should investigate the usability and utility of real-time interventions that detect and present deceptive patterns on the web. For example, such studies can evaluate different realizations of \name's extension that involve different trade-offs between real-time performance, usability, and utility. In such studies, participants can be prompted to install the extension, visit real-world websites, and answer surveys based on their experience.

\subsection{Developer Outreach}

To evaluate the effectiveness of the Lighthouse reports, we notified the Shopify web developers of the 4,492 website we analyzed in \Cref{subsec:measurements} about various deceptive patterns present on their websites.

\subsubsection{Outreach Design}
We crawled the \textit{Shopify Partners Directory} in search of developers and their most popular Shopify-based website. Next, we analyzed these websites for deceptive patterns as detailed earlier in \Cref{subsec:measurements}. Through that process we created Lighthouse~reports for each of the website. Next, we reached out to the web developers of the website through email, inquiring about~their opinion of the Lighthouse report. Particularly, we asked them 4 questions:
\begin{enumerate}[nosep]
    \item[Q1.] \textit{Was this report is useful to you?}
    \item[Q2.] \textit{Would you like more or less information included in the report, and if so, what?}
    \item[Q3.] \textit{In light of this report, would you be willing to make any changes?}
    \item[Q4.] \textit{Would you like to get access to this tool to run on your complete website?}
\end{enumerate}

\subsubsection{Ethical Consideration}
In our email to the developers, we clearly identified ourselves as developers and stated our purpose of creating a system to automatically create Lighthouse reports that identify deceptive patterns on the web, and that we wanted to understand their perspective on these reports. We also stated that no personally identifiable data was being collected and that their responses to our email were optional. Since, no PII data was collected for our study, the IRB at our institution certified our study as "not human subject research," and we were not required to obtain consent before sending out the emails.

\subsubsection{Findings}
Unfortunately, the study yielded only three usable replies. Two responses indicated that removing DPs was an owner-level decision for which they lacked authority, while the third cited the desire to maintain customer retention as the reason for not making changes.

\subsubsection{Future Studies} The limited response rate suggests that a different approach is needed to interact with developers. While prior work had success with email-based notifications for web developers~\cite{stover_how_2023}, those emails included language about vulnerabilities and legal liabilities. We opted not to use such language with the developers, which contributed to a reduced response rate. We suggest that future studies actively recruit developers and interview them about the Lighthouse reports. Another follow-up study can potentially help regulators generate variants of Lighthouse reports that check compliance with relevant laws, such as the French Data Protection Act.

}

\section{Limitations}
\label{sec:limitations}

\paragraph{Scope of Detectable Patterns}
A limitation for \name comes from the static UI analysis. Our approach inherently restricts the scope of detectable deceptive patterns to those visually present on a single page at a given time. Deceptive patterns such as \textit{nagging} cannot be detected through this approach, and thus, we filter Gary et. al.'s taxonomy~\cite{gray2023towards} to focus on static deceptive patterns.

\paragraph{Hardware Constraints}
Another practical limitation arises due to hardware limitations associated with deploying \name's extension using \texttt{T5}. The model requires approximately 1GB of memory, which might not be easily available on older devices. Furthermore, our experiments show that while near-real-time latency is achievable on modern hardware, the lack of GPU acceleration can significantly affect latency.

\paragraph{Multilingual Support}
\name's focus on English websites presents another limitation as the language module does not support non-English languages due to presence of language-specific datasets in the distillation pipeline. However, we note that it is possible to extend \name to other languages by using a multi-lingual language models, and curating a language specific distillation dataset.


\paragraph{YOLOv10 Ensemble Label Set}
Finally, the \textit{YOLOv10 Ensemble} is limited to identifying only a set of 7 common UI elements. Other interactive elements prevalent on modern websites, such as sliders and date pickers, are not explicitly recognized by the model. While this could potentially reduce the contextual information available to the \textit{Language Module}, we have empirically observed that deceptive patterns rarely use other interactive UI elements.

\section{Future Work}
\label{sec:future-work}

\name is a framework that takes a screenshot and returns localized deceptive patterns. We provide three sample applications that build on top of it. We envision that applications like \texttt{lighthouse-ci}~\cite{lighthouse-ci} can easily extend our work to integrate \name into developer CI/CD workflows. Regulators can also use \name with models aligned to their regulations to enforce policies automatically.

Another direction for future work is to enhance \name's multilingual capabilities. While deceptive patterns are language-agnostic, the current framework's training and evaluation datasets were focused on English-language websites. Future efforts could explore language-specific versions of the \textit{LanguageModule} or integrate advanced LLMs that can reason across multiple languages.

Lastly, we also envision using \name as a tool to generate large-scale datasets of websites with elements automatically labeled for deceptive patterns. A significant bottleneck for finetuning/retraining a VLLM for this task is the lack of large-scale annotated training data. Datasets created using \name can be utilized to overcome this challenge and potentially improve the performance of VLLMs in detecting deceptive patterns in the future.

\section{Conclusion}
\label{sec:conclusion}
In this paper, we introduce \name, a framework to automatically detect deceptive patterns on websites. \name employs a modular approach: first, it captures a screenshot of the website and processes it using the \textit{Vision Module} to provide a textual representation of the website (\textmap). It then analyzes the \textmap using a \textit{Language Module} to identify deceptive patterns and their type. We evaluate \name on a dataset of real-world websites to demonstrate its accuracy in identifying and localizing deceptive patterns. We then instantiate \name in three settings: a user-facing browser extension, a developer-facing Lighthouse report, and a researcher/regulator-facing website analysis tool.

\section*{Acknowledgments}
This work was supported by the NSF through awards CNS-1942014 and CNS-2247381, and by a research grant from the Google PSS Privacy Faculty Award program. Finally, we thank the reviewers for their thoughtful recommendations.


\bibliographystyle{abbrv}
\small{\bibliography{references.bib}}

@inproceedings{chen2023unveiling,
  title={Unveiling the tricks: Automated detection of dark patterns in mobile applications},
  author={Chen, Jieshan and Sun, Jiamou and Feng, Sidong and Xing, Zhenchang and Lu, Qinghua and Xu, Xiwei and Chen, Chunyang},
  booktitle={Proceedings of the 36th Annual ACM Symposium on User Interface Software and Technology},
  pages={1--20},
  year={2023}
}

@article{mathur2019dark,
  title={Dark patterns at scale: Findings from a crawl of 11K shopping websites},
  author={Mathur, Arunesh and Acar, Gunes and Friedman, Michael J and Lucherini, Eli and Mayer, Jonathan and Chetty, Marshini and Narayanan, Arvind},
  journal={Proceedings of the ACM on human-computer interaction},
  volume={3},
  number={CSCW},
  pages={1--32},
  year={2019},
  publisher={ACM New York, NY, USA}
}

@inproceedings{gray2023towards,
  title={Towards a preliminary ontology of dark patterns knowledge},
  author={Gray, Colin M and Santos, Cristiana and Bielova, Nataliia},
  booktitle={Extended abstracts of the 2023 CHI conference on human factors in computing systems},
  pages={1--9},
  year={2023}
}

@article{curley2021design,
  title={The Design of a framework for the detection of web-based dark patterns},
  author={Curley, Andrea and O'Sullivan, Dympna and Gordon, Damian and Tierney, Brendan and Stavrakakis, Ioannis},
  year={2021},
  publisher={Technological University Dublin}
}

@inproceedings{adorna2024developing,
  title={Developing a Browser Extension for the Automated Detection of Deceptive Patterns in Cookie Banners},
  author={Adorna, Juris Hannah and Dantis, Aurel Jared and Feria, Rommel and Figueroa, Ligaya Leah and Solamo, Rowena},
  booktitle={Proceedings of the Workshop on Computation: Theory and Practice (WCTP 2023)},
  volume={20},
  pages={101},
  year={2024},
  organization={Springer Nature}
}

@inproceedings{raju2022smart,
  title={Smart dark pattern detection: Making aware of misleading patterns through the intended app},
  author={Raju, S Hrushikesava and Waris, Saiyed Faiayaz and Adinarayna, S and Jadala, Vijaya Chandra and Rao, G Subba},
  booktitle={Sentimental Analysis and Deep Learning: Proceedings of ICSADL 2021},
  pages={933--947},
  year={2022},
  organization={Springer}
}

@inproceedings{gray2018dark,
  title={The dark (patterns) side of UX design},
  author={Gray, Colin M and Kou, Yubo and Battles, Bryan and Hoggatt, Joseph and Toombs, Austin L},
  booktitle={Proceedings of the 2018 CHI conference on human factors in computing systems},
  pages={1--14},
  year={2018}
}

@inproceedings{soe2020circumvention,
  title={Circumvention by design-dark patterns in cookie consent for online news outlets},
  author={Soe, Than Htut and Nordberg, Oda Elise and Guribye, Frode and Slavkovik, Marija},
  booktitle={Proceedings of the 11th nordic conference on human-computer interaction: Shaping experiences, shaping society},
  pages={1--12},
  year={2020}
}

@article{brown2020language,
  title={Language models are few-shot learners},
  author={Brown, Tom and Mann, Benjamin and Ryder, Nick and Subbiah, Melanie and Kaplan, Jared D and Dhariwal, Prafulla and Neelakantan, Arvind and Shyam, Pranav and Sastry, Girish and Askell, Amanda and others},
  journal={Advances in neural information processing systems},
  volume={33},
  pages={1877--1901},
  year={2020}
}

@misc{huggingface_transformersjs_webgpu,
  author       = {{Hugging Face}},
  title        = {{Running models on WebGPU}},
  year         = {2024},
  howpublished = {\url{https://huggingface.co/docs/transformers.js/en/guides/webgpu}},
  note         = {Accessed: 2025-04-13}
}

@misc{dark_pattern_site,
  title = {Deceptive Patterns},
  author = {Harry Brignull},
  howpublished = {\url{https://www.deceptive.design/}}
}

@inproceedings{zhang2021screen,
  title={Screen recognition: Creating accessibility metadata for mobile applications from pixels},
  author={Zhang, Xiaoyi and De Greef, Lilian and Swearngin, Amanda and White, Samuel and Murray, Kyle and Yu, Lisa and Shan, Qi and Nichols, Jeffrey and Wu, Jason and Fleizach, Chris and others},
  booktitle={Proceedings of the 2021 CHI Conference on Human Factors in Computing Systems},
  pages={1--15},
  year={2021}
}

@inproceedings{xie2020uied,
  title={UIED: a hybrid tool for GUI element detection},
  author={Xie, Mulong and Feng, Sidong and Xing, Zhenchang and Chen, Jieshan and Chen, Chunyang},
  booktitle={Proceedings of the 28th ACM Joint Meeting on European Software Engineering Conference and Symposium on the Foundations of Software Engineering},
  pages={1655--1659},
  year={2020}
}

@article{li2024ferret,
  title={Ferret-ui 2: Mastering universal user interface understanding across platforms},
  author={Li, Zhangheng and You, Keen and Zhang, Haotian and Feng, Di and Agrawal, Harsh and Li, Xiujun and Moorthy, Mohana Prasad Sathya and Nichols, Jeff and Yang, Yinfei and Gan, Zhe},
  journal={arXiv preprint arXiv:2410.18967},
  year={2024}
}

@article{lu2024omniparser,
  title={Omniparser for pure vision based gui agent},
  author={Lu, Yadong and Yang, Jianwei and Shen, Yelong and Awadallah, Ahmed},
  journal={arXiv preprint arXiv:2408.00203},
  year={2024}
}

@article{bosch2016tales,
  title={Tales from the dark side: Privacy dark strategies and privacy dark patterns},
  author={B{\"o}sch, Christoph and Erb, Benjamin and Kargl, Frank and Kopp, Henning and Pfattheicher, Stefan},
  journal={Proceedings on Privacy Enhancing Technologies},
  year={2016}
}

@inproceedings{mansur2023aidui,
  title={Aidui: Toward automated recognition of dark patterns in user interfaces},
  author={Mansur, SM Hasan and Salma, Sabiha and Awofisayo, Damilola and Moran, Kevin},
  booktitle={2023 IEEE/ACM 45th International Conference on Software Engineering (ICSE)},
  pages={1958--1970},
  year={2023},
  organization={IEEE}
}

@misc{GDPR2016,
  title        = {Regulation {(EU)} 2016/679 of the European Parliament and of the Council of 27 April 2016 on the protection of natural persons with regard to the processing of personal data and on the free movement of such data, and repealing Directive 95/46/EC (General Data Protection Regulation)},
  howpublished = {\url{https://eur-lex.europa.eu/eli/reg/2016/679/oj}},
  year         = {2016},
  note         = {Accessed: 2024-09-02}
}

@online{Insite2023,
  author = {Wang, David and Anthony Ribando, John and Mo, David and Tung, Nicholas},
  title = {insite. A chrome extension that protects consumers from marketing tricks when they shop online.},
  year = 2019,
  url = {https://devpost.com/software/insite-qfpjcd},
  lastaccessed = {6 Sep 2022}
}

@inproceedings{sermuga2021uisketch,
  title={UISketch: a large-scale dataset of UI element sketches},
  author={Sermuga Pandian, Vinoth Pandian and Suleri, Sarah and Jarke, Prof Dr Matthias},
  booktitle={Proceedings of the 2021 CHI Conference on Human Factors in Computing Systems},
  pages={1--14},
  year={2021}
}

@misc{extcolors,
  author       = {Thomas Cairns},
  title        = {extcolors: Extract colors from an image using k-means clustering},
  howpublished = {\url{https://pypi.org/project/extcolors/}},
  year         = {2021},
  note         = {Accessed: 2024-09-02}
}

@article{hsieh2023distilling,
  title={Distilling step-by-step! outperforming larger language models with less training data and smaller model sizes},
  author={Hsieh, Cheng-Yu and Li, Chun-Liang and Yeh, Chih-Kuan and Nakhost, Hootan and Fujii, Yasuhisa and Ratner, Alexander and Krishna, Ranjay and Lee, Chen-Yu and Pfister, Tomas},
  journal={arXiv preprint arXiv:2305.02301},
  year={2023}
}

@article{ouyang2022training,
  title={Training language models to follow instructions with human feedback},
  author={Ouyang, Long and Wu, Jeffrey and Jiang, Xu and Almeida, Diogo and Wainwright, Carroll and Mishkin, Pamela and Zhang, Chong and Agarwal, Sandhini and Slama, Katarina and Ray, Alex and others},
  journal={Advances in neural information processing systems},
  volume={35},
  pages={27730--27744},
  year={2022}
}

@article{shool2025systematic,
  title={A systematic review of large language model (LLM) evaluations in clinical medicine},
  author={Shool, Sina and Adimi, Sara and Saboori Amleshi, Reza and Bitaraf, Ehsan and Golpira, Reza and Tara, Mahmood},
  journal={BMC Medical Informatics and Decision Making},
  volume={25},
  number={1},
  pages={117},
  year={2025},
  publisher={Springer}
}

@article{agarwal2024llm,
  title={" Which LLM should I use?": Evaluating LLMs for tasks performed by Undergraduate Computer Science Students in India},
  author={Agarwal, Vibhor and Thureja, Nakul and Krishan Garg, Madhav and Dharmavaram, Sahiti and Kumar, Dhruv and others},
  journal={arXiv e-prints},
  pages={arXiv--2402},
  year={2024}
}

@inproceedings{renze2024effect,
  title={The effect of sampling temperature on problem solving in large language models},
  author={Renze, Matthew},
  booktitle={Findings of the Association for Computational Linguistics: EMNLP 2024},
  pages={7346--7356},
  year={2024}
}

@article{cui2025curie,
  title={CURIE: Evaluating LLMs On Multitask Scientific Long Context Understanding and Reasoning},
  author={Cui, Hao and Shamsi, Zahra and Cheon, Gowoon and Ma, Xuejian and Li, Shutong and Tikhanovskaya, Maria and Norgaard, Peter and Mudur, Nayantara and Plomecka, Martyna and Raccuglia, Paul and others},
  journal={arXiv preprint arXiv:2503.13517},
  year={2025}
}

@article{zhou2023instruction,
  title={Instruction-following evaluation for large language models},
  author={Zhou, Jeffrey and Lu, Tianjian and Mishra, Swaroop and Brahma, Siddhartha and Basu, Sujoy and Luan, Yi and Zhou, Denny and Hou, Le},
  journal={arXiv preprint arXiv:2311.07911},
  year={2023}
}

@article{achiam2023gpt,
  title={Gpt-4 technical report},
  author={Achiam, Josh and Adler, Steven and Agarwal, Sandhini and Ahmad, Lama and Akkaya, Ilge and Aleman, Florencia Leoni and Almeida, Diogo and Altenschmidt, Janko and Altman, Sam and Anadkat, Shyamal and others},
  journal={arXiv preprint arXiv:2303.08774},
  year={2023}
}

@article{pochat2018tranco,
  title={Tranco: A research-oriented top sites ranking hardened against manipulation},
  author={Pochat, Victor Le and Van Goethem, Tom and Tajalizadehkhoob, Samaneh and Korczy{\'n}ski, Maciej and Joosen, Wouter},
  journal={arXiv preprint arXiv:1806.01156},
  year={2018}
}

@article{wang2024yolov10,
  title={Yolov10: Real-time end-to-end object detection},
  author={Wang, Ao and Chen, Hui and Liu, Lihao and Chen, Kai and Lin, Zijia and Han, Jungong and Ding, Guiguang},
  journal={arXiv preprint arXiv:2405.14458},
  year={2024}
}

@article{ren2015faster,
  title={Faster r-cnn: Towards real-time object detection with region proposal networks},
  author={Ren, Shaoqing},
  journal={arXiv preprint arXiv:1506.01497},
  year={2015}
}

@article{redmon2018yolov3,
  title={Yolov3: An incremental improvement},
  author={Redmon, Joseph},
  journal={arXiv preprint arXiv:1804.02767},
  year={2018}
}

@article{wei2022chain,
  title={Chain-of-thought prompting elicits reasoning in large language models},
  author={Wei, Jason and Wang, Xuezhi and Schuurmans, Dale and Bosma, Maarten and Xia, Fei and Chi, Ed and Le, Quoc V and Zhou, Denny and others},
  journal={Advances in neural information processing systems},
  volume={35},
  pages={24824--24837},
  year={2022}
}

@inproceedings{nayak2024experimental,
  title={Experimental Security Analysis of Sensitive Data Access by Browser Extensions},
  author={Nayak, Asmit and Khandelwal, Rishabh and Fernandes, Earlence and Fawaz, Kassem},
  booktitle={Proceedings of the ACM on Web Conference 2024},
  pages={1283--1294},
  year={2024}
}

@misc{SchneiderWallace2023,
  author       = {{Schneider Wallace}},
  title        = {Dark Patterns: Making Online Subscriptions Harder to Cancel Draws Lawsuits, Settlements, and Government Scrutiny},
  howpublished ={\url{https://www.schneiderwallace.com/media/dark-patterns-making-online-subscriptions-harder-to-cancel-draws-lawsuits-settlements-and-government-scrutiny/}},
  year         = {2023},
}

@misc{Jika2023,
  author       = {Spike aka Steve Spiker},
  title        = {Tweet by Spike aka Steve Spiker on X},
  howpublished = {\url{https://x.com/spjika/status/1686492710910427137}},
  year         = {2023},
  note         = {Accessed: 2024-09-02}
}

@misc{luke2019,
  author       = {Luke Stein},
  title        = {Tweet by Luke Stein on X},
  howpublished = {\url{https://x.com/lukestein/status/1150014732486742016}},
  year         = {2019},
  note         = {Accessed: 2024-09-02}
}

@inproceedings{conti2010malicious,
  title={Malicious interface design: exploiting the user},
  author={Conti, Gregory and Sobiesk, Edward},
  booktitle={Proceedings of the 19th international conference on World wide web},
  pages={271--280},
  year={2010}
}

@inproceedings{khandelwal_prisec_2021,
	title = {\{{PriSEC}\}: {A} {Privacy} {Settings} {Enforcement} {Controller}},
	isbn = {978-1-939133-24-3},
	shorttitle = {\{{PriSEC}\}},
	url = {https://www.usenix.org/conference/usenixsecurity21/presentation/khandelwal},
	language = {en},
	urldate = {2024-09-03},
	author = {Khandelwal, Rishabh and Linden, Thomas and Harkous, Hamza and Fawaz, Kassem},
	year = {2021},
	pages = {465--482},
}

@article{wu2023webui, 
    title={WebUI: A Dataset for Enhancing Visual UI Understanding with Web Semantics}, 
    author={Jason Wu and Siyan Wang and Siman Shen and Yi-Hao Peng and Jeffrey Nichols and Jeffrey Bigham}, 
    journal={ACM Conference on Human Factors in Computing Systems (CHI)}, 
    year={2023}
}

@article{brooke1996sus,
  title={SUS: A quick and dirty usability scale},
  author={Brooke, J},
  journal={Usability Evaluation in INdustry/Taylor and Francis},
  year={1996}
}

@inproceedings{ueid_xie,
author = {Xie, Mulong and Feng, Sidong and Xing, Zhenchang and Chen, Jieshan and Chen, Chunyang},
title = {UIED: a hybrid tool for GUI element detection},
year = {2020},
isbn = {9781450370431},
publisher = {Association for Computing Machinery},
address = {New York, NY, USA},
url = {https://doi.org/10.1145/3368089.3417940},
doi = {10.1145/3368089.3417940},
booktitle = {Proceedings of the 28th ACM Joint Meeting on European Software Engineering Conference and Symposium on the Foundations of Software Engineering},
pages = {1655–1659},
numpages = {5},
location = {Virtual Event, USA},
series = {ESEC/FSE 2020}
}

@inproceedings{rico,
 author = {Liu, Thomas F. and Craft, Mark and Situ, Jason and Yumer, Ersin and Mech, Radomir and Kumar, Ranjitha},
 title = {Learning Design Semantics for Mobile Apps},
 booktitle = {The 31st Annual ACM Symposium on User Interface Software and Technology},
 series = {UIST '18},
 year = {2018},
 isbn = {978-1-4503-5948-1},
 location = {Berlin, Germany},
 pages = {569--579},
 numpages = {11},
 url = {http://doi.acm.org/10.1145/3242587.3242650},
 doi = {10.1145/3242587.3242650},
 acmid = {3242650},
 address = {New York, NY, USA},
}

@article{khandelwal_cookie,
	title = {Automated {Cookie} {Notice} {Analysis} and {Enforcement}},
	language = {en},
	author = {Khandelwal, Rishabh and Nayak, Asmit and Harkous, Hamza and Fawaz, Kassem},
}

@incollection{lewis_gameful_2014,
	address = {Berkeley, CA},
	title = {Gameful {Patterns}},
	isbn = {978-1-4302-6422-4},
	url = {https://doi.org/10.1007/978-1-4302-6422-4_4},
	language = {en},
	urldate = {2024-09-04},
	booktitle = {Irresistible {Apps}: {Motivational} {Design} {Patterns} for {Apps}, {Games}, and {Web}-based {Communities}},
	publisher = {Apress},
	author = {Lewis, Chris},
	editor = {Lewis, Chris},
	year = {2014},
	doi = {10.1007/978-1-4302-6422-4_4},
	pages = {33--50},
}

@inproceedings{bannihatti_kumar_finding_2020,
	address = {Taipei Taiwan},
	title = {Finding a {Choice} in a {Haystack}: {Automatic} {Extraction} of {Opt}-{Out} {Statements} from {Privacy} {Policy} {Text}},
	isbn = {978-1-4503-7023-3},
	shorttitle = {Finding a {Choice} in a {Haystack}},
	url = {https://dl.acm.org/doi/10.1145/3366423.3380262},
	doi = {10.1145/3366423.3380262},
	language = {en},
	urldate = {2024-09-04},
	booktitle = {Proceedings of {The} {Web} {Conference} 2020},
	publisher = {ACM},
	author = {Bannihatti Kumar, Vinayshekhar and Iyengar, Roger and Nisal, Namita and Feng, Yuanyuan and Habib, Hana and Story, Peter and Cherivirala, Sushain and Hagan, Margaret and Cranor, Lorrie and Wilson, Shomir and Schaub, Florian and Sadeh, Norman},
	month = apr,
	year = {2020},
	pages = {1943--1954},
}

@misc{CPRA,
  author       = {{California Privacy Protection Agency}},
  title        = {{California Privacy Rights Act (CPRA)}},
  howpublished = {\url{https://thecpra.org/}},
  note         = {Accessed: 2024-09-04},
  year         = {2024}
}

@article{landis1977measurement,
  title={The measurement of observer agreement for categorical data},
  author={Landis, J Richard and Koch, Gary G},
  journal={biometrics},
  pages={159--174},
  year={1977},
  publisher={JSTOR}
}

@article{zhang2024visually,
  title={Why are visually-grounded language models bad at image classification?},
  author={Zhang, Yuhui and Unell, Alyssa and Wang, Xiaohan and Ghosh, Dhruba and Su, Yuchang and Schmidt, Ludwig and Yeung-Levy, Serena},
  journal={arXiv preprint arXiv:2405.18415},
  year={2024}
}

@online{lighthouse_overview,
    author = "{Google Chrome Developers}",
    title = "{Lighthouse Overview}",
    url = "[https://developer.chrome.com/docs/lighthouse/overview](https://developer.chrome.com/docs/lighthouse/overview)",
    urldate = {2025-03-10}
}

@article{GoogleGemini,
  title={Gemini: a family of highly capable multimodal models},
  author={Team, Gemini and Anil, Rohan and Borgeaud, Sebastian and Alayrac, Jean-Baptiste and Yu, Jiahui and Soricut, Radu and Schalkwyk, Johan and Dai, Andrew M and Hauth, Anja and Millican, Katie and others},
  journal={arXiv preprint arXiv:2312.11805},
  year={2023}
}

@misc{qwen2.5,
    title = {Qwen2.5: A Party of Foundation Models},
    url = {https://qwenlm.github.io/blog/qwen2.5/},
    author = {Qwen Team},
    month = {September},
    year = {2024}
}

@misc{langdetect,
  author       = {Nakatani Shuyo and others (ported to Python)},
  title        = {langdetect: Language detection library ported to Python},
  howpublished = {\url{https://pypi.org/project/langdetect/}},
  note         = {Version [Specify version if needed, e.g., 1.0.9]},
  year         = {2021}
}

@misc{nudenet,
  author       = {notAI-tech},
  title        = {NudeNet: Nudity detection with deep neural networks},
  howpublished = {\url{https://github.com/notAI-tech/NudeNet}},
  year         = {2019},
  note         = {Accessed March 19, 2025}
}

@misc{lighthouse_chrome,
	title = {Lighthouse - {Chrome} {Web} {Store}},
	url = {https://chromewebstore.google.com/detail/lighthouse/blipmdconlkpinefehnmjammfjpmpbjk},
	language = {en},
	urldate = {2025-03-24},
}

@article{stover_how_2023,
	title = {How {Website} {Owners} {Face} {Privacy} {Issues}: {Thematic} {Analysis} of {Responses} from a {Covert} {Notification} {Study} {Reveals} {Diverse} {Circumstances} and {Challenges}},
	volume = {2023},
	copyright = {https://creativecommons.org/licenses/by/4.0/},
	issn = {2299-0984},
	shorttitle = {How {Website} {Owners} {Face} {Privacy} {Issues}},
	url = {https://petsymposium.org/popets/2023/popets-2023-0051.php},
	doi = {10.56553/popets-2023-0051},
	language = {en},
	number = {2},
	urldate = {2025-03-24},
	journal = {Proceedings on Privacy Enhancing Technologies},
	author = {Stöver, Alina and Gerber, Nina and Pridöhl, Henning and Maass, Max and Bretthauer, Sebastian and Spiecker Gen. Döhmann, Indra and Hollick, Matthias and Herrmann, Dominik},
	month = apr,
	year = {2023},
	pages = {251--264},
}

@misc{lighthouse_gatherer,
author = {Google},
  title = {full-page-screenshot.js},
  publisher = {GitHub},
  journal = {GitHub repository},
  howpublished = {\url{https://github.com/GoogleChrome/lighthouse/blob/main/core/gather/gatherers/full-page-screenshot.js}},
}

@misc{lighthouse-ci,
author = {Google},
  title = {Lighthouse-ci},
  publisher = {GitHub},
  journal = {GitHub repository},
  howpublished = {\url{https://github.com/GoogleChrome/lighthouse-ci}},
}

@misc{deepscaler2025,
  title={DeepScaleR: Surpassing O1-Preview with a 1.5B Model by Scaling RL},
  author={Michael Luo and Sijun Tan and Justin Wong and Xiaoxiang Shi and William Y. Tang and Manan Roongta and Colin Cai and Jeffrey Luo and Tianjun Zhang and Li Erran Li and Raluca Ada Popa and Ion Stoica},
  year={2025},
  howpublished={\url{https://pretty-radio-b75.notion.site/DeepScaleR-Surpassing-O1-Preview-with-a-1-5B-Model-by-Scaling-RL-19681902c1468005bed8ca303013a4e2}},
  note={Notion Blog}
}

@article{FlanT5,
  title={Scaling instruction-finetuned language models},
  author={Chung, Hyung Won and Hou, Le and Longpre, Shayne and Zoph, Barret and Tay, Yi and Fedus, William and Li, Yunxuan and Wang, Xuezhi and Dehghani, Mostafa and Brahma, Siddhartha and others},
  journal={Journal of Machine Learning Research},
  volume={25},
  number={70},
  pages={1--53},
  year={2024}
}

@article{T5,
  title={Exploring the limits of transfer learning with a unified text-to-text transformer},
  author={Roberts, Adam and Raffel, Colin and Lee, Katherine and Matena, Michael and Shazeer, Noam and Liu, Peter J and Narang, Sharan and Li, Wei and Zhou, Yanqi},
  journal={Google Research},
  year={2019}
}

@inproceedings{50Shades,
  title={50 Shades of Deceptive Patterns: A Unified Taxonomy, Multimodal Detection, and Security Implications},
  author={Shi, Zewei and Sun, Ruoxi and Chen, Jieshan and Sun, Jiamou and Xue, Minhui and Gao, Yansong and Liu, Feng and Yuan, Xingliang},
  booktitle={Proceedings of the ACM Web Conference 2025 (WWW'25)},
  year={2025}
}

@article{chen2023shikra,
  title={Shikra: Unleashing multimodal llm's referential dialogue magic},
  author={Chen, Keqin and Zhang, Zhao and Zeng, Weili and Zhang, Richong and Zhu, Feng and Zhao, Rui},
  journal={arXiv preprint arXiv:2306.15195},
  year={2023}
}

@article{you2023ferret,
  title={Ferret: Refer and ground anything anywhere at any granularity},
  author={You, Haoxuan and Zhang, Haotian and Gan, Zhe and Du, Xianzhi and Zhang, Bowen and Wang, Zirui and Cao, Liangliang and Chang, Shih-Fu and Yang, Yinfei},
  journal={arXiv preprint arXiv:2310.07704},
  year={2023}
}

@article{li2025towards,
  title={Towards Visual Text Grounding of Multimodal Large Language Model},
  author={Li, Ming and Zhang, Ruiyi and Chen, Jian and Gu, Jiuxiang and Zhou, Yufan and Dernoncourt, Franck and Zhu, Wanrong and Zhou, Tianyi and Sun, Tong},
  journal={arXiv preprint arXiv:2504.04974},
  year={2025}
}

@article{shiri2024empirical,
  title={An Empirical Analysis on Spatial Reasoning Capabilities of Large Multimodal Models},
  author={Shiri, Fatemeh and Guo, Xiao-Yu and Far, Mona Golestan and Yu, Xin and Haffari, Gholamreza and Li, Yuan-Fang},
  journal={arXiv preprint arXiv:2411.06048},
  year={2024}
}

@misc{openai_learning_reason_2024,
  title = {Learning to reason with {LLMs}},
  author = {{OpenAI}},
  howpublished = {\url{https://openai.com/index/learning-to-reason-with-llms/}},
  url = {https://openai.com/index/learning-to-reason-with-llms/},  
}

@article{guo2025deepseek,
  title={Deepseek-r1: Incentivizing reasoning capability in llms via reinforcement learning},
  author={Guo, Daya and Yang, Dejian and Zhang, Haowei and Song, Junxiao and Zhang, Ruoyu and Xu, Runxin and Zhu, Qihao and Ma, Shirong and Wang, Peiyi and Bi, Xiao and others},
  journal={arXiv preprint arXiv:2501.12948},
  year={2025}
}

@article{zang2025contextual,
  title={Contextual object detection with multimodal large language models},
  author={Zang, Yuhang and Li, Wei and Han, Jun and Zhou, Kaiyang and Loy, Chen Change},
  journal={International Journal of Computer Vision},
  volume={133},
  number={2},
  pages={825--843},
  year={2025},
  publisher={Springer}
}

@article{molmo,
  title={Molmo and pixmo: Open weights and open data for state-of-the-art multimodal models},
  author={Deitke, Matt and Clark, Christopher and Lee, Sangho and Tripathi, Rohun and Yang, Yue and Park, Jae Sung and Salehi, Mohammadreza and Muennighoff, Niklas and Lo, Kyle and Soldaini, Luca and others},
  journal={arXiv preprint arXiv:2409.17146},
  year={2024}
}

@article{leevy2018survey,
  title={A survey on addressing high-class imbalance in big data},
  author={Leevy, Joffrey L and Khoshgoftaar, Taghi M and Bauder, Richard A and Seliya, Naeem},
  journal={Journal of Big Data},
  volume={5},
  number={1},
  pages={1--30},
  year={2018},
  publisher={Springer}
}

@article{elsoud2024under,
  title={Under Sampling Techniques for Handling Unbalanced Data with Various Imbalance Rates: A Comparative Study.},
  author={Elsoud, Esraa Abu and Hassan, Mohamad and Alidmat, Omar and Al Henawi, Esraa and Alshdaifat, Nawaf and Igtait, Mosab and Ghaben, Ayman and Katrawi, Anwar and Dmour, Mohmmad},
  journal={International Journal of Advanced Computer Science \& Applications},
  volume={15},
  number={8},
  year={2024}
}

@misc{wu_never-ending_2023,
	title = {Never-ending {Learning} of {User} {Interfaces}},
	url = {http://arxiv.org/abs/2308.08726},
	doi = {10.48550/arXiv.2308.08726},
	language = {en},
	urldate = {2025-04-13},
	publisher = {arXiv},
	author = {Wu, Jason and Krosnick, Rebecca and Schoop, Eldon and Swearngin, Amanda and Bigham, Jeffrey P. and Nichols, Jeffrey},
	month = aug,
	year = {2023},
	note = {arXiv:2308.08726 [cs]}
}

@misc{shadcn,
	title = {Build your component library - shadcn/ui},
	url = {https://ui.shadcn.com/},
	language = {en},
	urldate = {2025-04-13},
	author = {shadcn}
}

@article{mohankumar_benchmark_nodate,
	title = {A benchmark dataset and ensemble {YOLO} method for enhanced underwater fish detection},
	volume = {n/a},
	copyright = {1225-6463/\$ © 2025 ETRI},
	issn = {2233-7326},
	url = {https://onlinelibrary.wiley.com/doi/abs/10.4218/etrij.2024-0383},
	doi = {10.4218/etrij.2024-0383},
	language = {en},
	number = {n/a},
	urldate = {2025-04-14},
	journal = {ETRI Journal},
	author = {Mohankumar, Vijayalakshmi and Anbalagan, Sasithradevi},
	note = {\_eprint: https://onlinelibrary.wiley.com/doi/pdf/10.4218/etrij.2024-0383},
}

@article{walambe_lightweight_2021,
	title = {Lightweight {Object} {Detection} {Ensemble} {Framework} for {Autonomous} {Vehicles} in {Challenging} {Weather} {Conditions}},
	volume = {2021},
	issn = {1687-5265},
	url = {https://www.ncbi.nlm.nih.gov/pmc/articles/PMC8516532/},
	doi = {10.1155/2021/5278820},
	urldate = {2025-04-14},
	journal = {Computational Intelligence and Neuroscience},
	author = {Walambe, Rahee and Marathe, Aboli and Kotecha, Ketan and Ghinea, George},
	month = oct,
	year = {2021},
	pmid = {34659392},
	pmcid = {PMC8516532},
	pages = {5278820},
}

@misc{pham_optimizing_2024,
	title = {Optimizing {YOLO} {Architectures} for {Optimal} {Road} {Damage} {Detection} and {Classification}: {A} {Comparative} {Study} from {YOLOv7} to {YOLOv10}},
	shorttitle = {Optimizing {YOLO} {Architectures} for {Optimal} {Road} {Damage} {Detection} and {Classification}},
	url = {http://arxiv.org/abs/2410.08409},
	doi = {10.48550/arXiv.2410.08409},
	urldate = {2025-04-14},
	publisher = {arXiv},
	author = {Pham, Vung and Ngoc, Lan Dong Thi and Bui, Duy-Linh},
	month = oct,
	year = {2024},
	note = {arXiv:2410.08409 [cs]},
}

@misc{Label_Studio,
  title={{Label Studio}: Data labeling software},
  url={https://github.com/HumanSignal/label-studio},
  note={Open source software available from https://github.com/HumanSignal/label-studio},
  author={
    Maxim Tkachenko and
    Mikhail Malyuk and
    Andrey Holmanyuk and
    Nikolai Liubimov},
  year={2020-2025},
}

@misc{api_security,
	title = {The 2023 {State} of {API} {Security} {Report} - {Global} {Findings}},
	url = {https://www.traceable.ai/2023-state-of-api-security},
	language = {en-US},
	urldate = {2025-04-14},
	month = apr,
	year = {2023},
}

@misc{brignull-2023,
author = {Brignull, H and Leiser, M and Santos, C and Doshi, K},
month = {4},
title = {{Deceptive patterns – user interfaces designed to trick you}},
year = {2023},
url = {https://www.deceptive.design/},
}

@misc{AnthropicHaikuFineTune2024,
  author       = {Anthropic},
  title        = {Fine-tune Claude 3 Haiku in Amazon Bedrock},
  year         = {2024},
  month        = {July},
  url          = {https://www.anthropic.com/news/fine-tune-claude-3-haiku},
  howpublished = {\url{https://www.anthropic.com/news/fine-tune-claude-3-haiku}},
  note         = {Accessed: April 15, 2025}
}

@software{zotero,
  author       = {{Corporation for Digital Scholarship}},
  title        = {Zotero},
  url          = {https://www.zotero.org/},
    howpublished = {\url{https://www.zotero.org/}},
  version      = {7.0.0}, 
  date         = {2024}, 
  publisher    = {Corporation for Digital Scholarship},
  address      = {Vienna, VA}
}

@standard{w3c_webgpu_2024,
  title        = {{WebGPU}},
  organization = {{World Wide Web Consortium (W3C)}},
  type         = {W3C Recommendation},
  year         = {2024},
  month        = feb,
  day          = {13},
  url          = {https://www.w3.org/TR/webgpu/},
  urldate      = {2025-04-15},
  note         = {Defines a modern graphics and compute API for the Web.}
}

@article{leiser2023dark,
  title={Dark Patterns, Enforcement, and the emerging Digital Design Acquis: Manipulation beneath the Interface},
  author={Leiser, Mark and Santos, Cristiana},
  year={2023}
}

@inproceedings{sazid2023automated,
  title={Automated Detection of Dark Patterns Using In-Context Learning Capabilities of GPT-3},
  author={Sazid, Yasin and Fuad, Mridha Md Nafis and Sakib, Kazi},
  booktitle={2023 30th Asia-Pacific Software Engineering Conference (APSEC)},
  pages={569--573},
  year={2023},
  organization={IEEE}
}

@inproceedings{schafer2025don,
  title={Don't Detect, Just Correct: Can LLMs Defuse Deceptive Patterns Directly?},
  author={Sch{\"a}fer, Ren{\'e} and Preuschoff, Paul Miles and Niewianda, Rene and Hahn, Sophie and Fiedler, Kevin and Borchers, Jan},
  booktitle={Proceedings of the Extended Abstracts of the CHI Conference on Human Factors in Computing Systems},
  pages={1--11},
  year={2025}
}

@inproceedings{tan2025htmlrag,
  title={Htmlrag: Html is better than plain text for modeling retrieved knowledge in rag systems},
  author={Tan, Jiejun and Dou, Zhicheng and Wang, Wen and Wang, Mang and Chen, Weipeng and Wen, Ji-Rong},
  booktitle={Proceedings of the ACM on Web Conference 2025},
  pages={1733--1746},
  year={2025}
}

\clearpage
\begin{flushleft}
\section*{Appendix}
\end{flushleft}
\label{sec:appendix}

\subsection{Full Taxonomy}
\label{app:taxonomy}

\vspace{1em}
\begin{minipage}{\textwidth}
\captionof{table}{Filtered taxonomy of Deceptive Patterns. The table shows the Category, the Sub-Type, the description of the deceptive pattern, and an example.}
\centering
\resizebox{0.92\textwidth}{!}{
\begin{tabularx}{\textwidth}{m{1.7cm} 
>{\raggedright\arraybackslash}m{2.3cm} 
>{\raggedright\arraybackslash}X 
>{\raggedright\arraybackslash}X}
 \textbf{Category}  & \textbf{Sub-Type} & \textbf{Description} & \textbf{Examples} \\
\toprule
    \multirow{6}{=}{\textbf{Interface\\Interference}} & Confirmshaming & Guilt-tripping users into making a specific choice & ``No, I prefer to pay more''  \\
    \cmidrule(lr){2-4}
    & Fake-Scarcity/ Fake-Urgency & Create a false sense of urgency/scarcity to pressure users into making a choice & ``Only 3 left in stock'' \\
    \cmidrule(lr){2-4}
    & Nudge & Nudge a user towards a specific choice. & ``Accept All'' in bright colors, while ``Reject'' is hard to notice \\
    \midrule
    \textbf{Forced-Action} & Forced-Action & Design tactic forcing users to complete a specific task to proceed. & Pop-up ads \\
    \midrule

    \multirow{6}{3cm}{\textbf{Obstruction}} &  Pre-Selection & Choices given to users are already selected. & The checkbox of ``Sign up for news and updates'' is checked by default. \\
    \cmidrule(lr){2-4}
  & Visual Interference & Misleading design elements that distract or mislead users from important information. &  The term of use for service users are signing up for is in tiny font and cannot be clearly seen on sites.\\
    \cmidrule(lr){2-4}
  & Jargon & The use of non-user-friendly language to prevent users from understanding important information. & ``By affirming this selection, you consent to the perpetuation of automatic pecuniary transactions at designated intervals.'' \\
  \midrule

    \multirow{6}{3cm}{\textbf{Sneaking}} & Hidden Subscription & Users are not clearly informed that they are signing up for a service. & ``By signing up for this email, you are agreeing to news and information from us'' \\
    \cmidrule(lr){2-4}
  & Hidden Costs & Users are not clearly informed about all the costs associated with a service. & During checkout, unexpected fees such as ``handling charges''appear. \\
    \cmidrule(lr){2-4}
  & Disguised Ads & Visually misleading ads embedded into page content. & Prominent ``Download'' button at the top of a page that redirects the user to unrelated adware. \\
    \cmidrule(lr){2-4}
  & Trickwording & The use of non-user-friendly language to trick users into making certain choices & ``Newsletter subscription by default, tick here to unsubscribe'' \\
    \midrule
\textbf{Non-Deceptive} & & Common, user-friendly design element that does not show any deceptive pattern. & Any text or web element that does not exhibit any deceptive behavior. \\
\bottomrule
\end{tabularx}}
\label{tab:taxonomy-full}
\end{minipage}

\FloatBarrier

\subsection{Website with Multiple Deceptive Pattern}
\label{app:bbb}
\begin{figure}[h]
    \centering
    \includegraphics[width=\columnwidth]{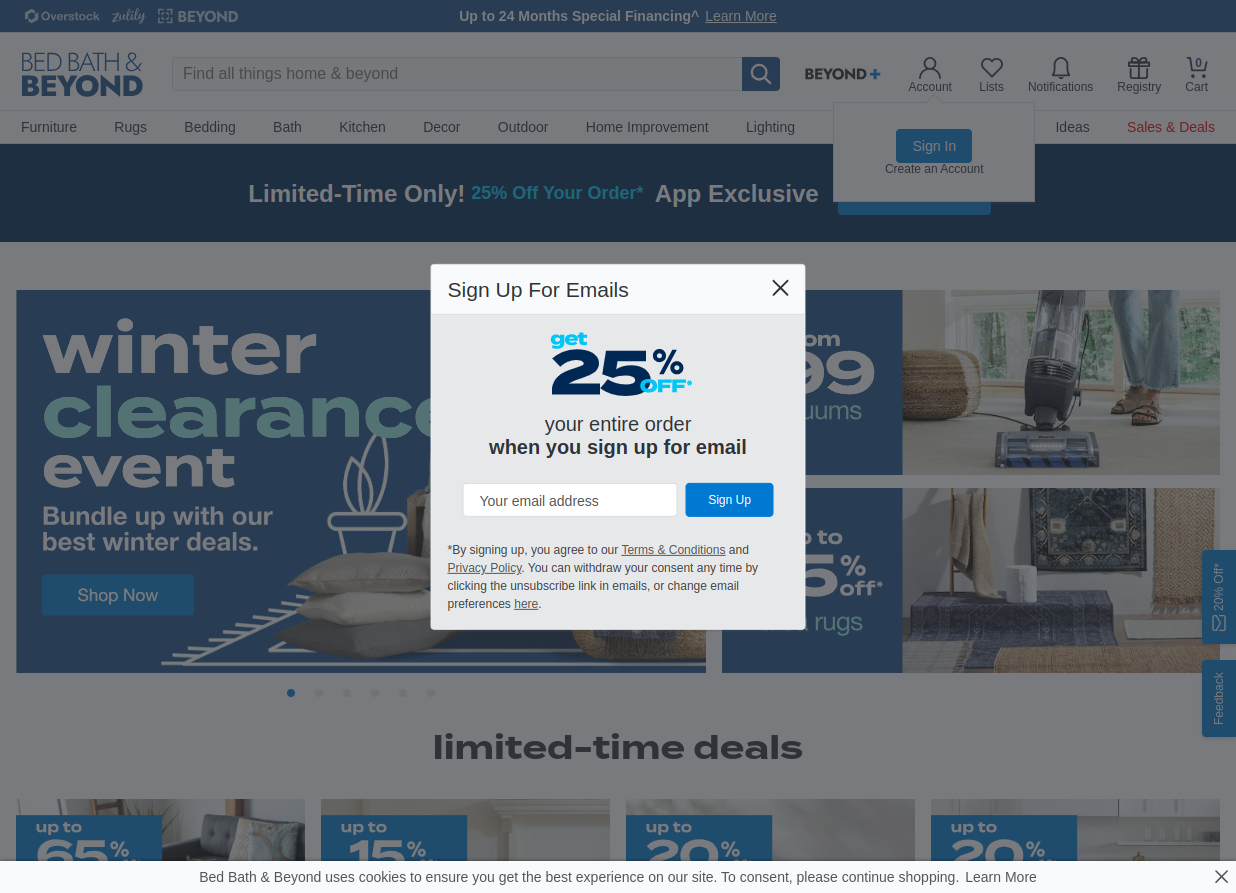}
    \caption{The screenshot of Bed, Bath, and Beyond, is identified as having four deceptive pattern categories on its web page.}
\end{figure}

\clearpage
\begin{flushleft}
\subsection{Mapping to Gary et al.'s Taxonomy}
\label{app:mapping}
\end{flushleft}




\begin{table}[ht]
    \centering
    \caption{Mapping taxonomy of Deceptive Patterns to Gray et al.~\cite{gray2023towards}}
    \label{tab:taxonomy-mapping}
    \resizebox{\columnwidth}{!}{
    \begin{tabular}{p{4cm}ll}
    \toprule
    \textbf{Category} & \textbf{Sub-Type} & \textbf{Mapping} \\
    \midrule
    \multirow{3}{*}{\textbf{\revision{Interface Interference}}} & \revision{Confirmshaming} & Social Engineering; Personalization \\
    \cmidrule(lr){2-3}
    & \revision{Fake-Scarcity / Fake-Urgency} & Social Engineering; Urgency\\
    \cmidrule(lr){2-3}
    & \revision{Nudge} & Interface Interference; Manipulating Visual Choice Architectures\\
    \midrule
    \textbf{\revision{Forced Action}} & \revision{Forced-Action} & Forced Action\\
    \midrule
    \multirow{3}{*}{\textbf{\revision{Obstruction}}} & \revision{Pre-Selection} & Interface Interference; Bad Default \\
    \cmidrule(lr){2-3}
    & \revision{Visual Interference} & Obstruction; Creating Barriers\\
    \cmidrule(lr){2-3}
    & \revision{Jargon} & Obstruction; Creating Barriers\\
    \midrule
    \multirow{4}{*}{\textbf{\revision{Sneaking}}} & \revision{Hidden Subscription} & Sneaking; Hiding Information\\
    \cmidrule(lr){2-3}
    & \revision{Hidden Costs} & Sneaking; Hiding Information\\
    \cmidrule(lr){2-3}
    & \revision{Disguised Ads} & Sneaking; Bait and Switch \\
    \cmidrule(lr){2-3}
    & \revision{Trick Wording} & Sneaking; (De)contextualizing Cues\\
    \bottomrule
    \end{tabular}
    }
\end{table}


\subsection{Website Generation Prompts}
\label{app:website-generation}
To generate websites using \texttt{v0} we used prompts from GPT4. Some example prompts are shown below:

\begin{enumerate}
    \item Create a virtual learning platform with courses, webinars, and interactive tools for students of all ages.
    \item Create an online gourmet food shop featuring high-quality ingredients, kitchen tools, and gourmet recipes.
    \item Design a warm and inviting UI for a pet care blog that radiates friendliness and approachability. The primary color palette should include soft, earthy tones with playful accents. Use a clean, easy-to-read sans serif font and incorporate elements like paw prints or pet silhouettes to enhance the thematic appeal. The homepage must prominently feature an engaging welcome message, and sections for various pets like dogs, cats, birds, and more, encouraging user navigation. Include a dynamic sidebar with widgets for pet care tips, a search bar, and featured posts. Dropdown menus should be intuitive, providing categories such as nutrition, training, health, and grooming. Visuals are key: integrate heart-warming pet images and infographics to explain care practices. End with footer links to contact details, social media, and a cute, animated pet mascot offering useful tips periodically.
\end{enumerate}

\subsection{Latency on Machines}
\label{app:latency}
\begin{figure}[h]
    \centering
    \begin{subfigure}[t]{0.49\columnwidth}
        \centering
        \includegraphics[width=\textwidth]{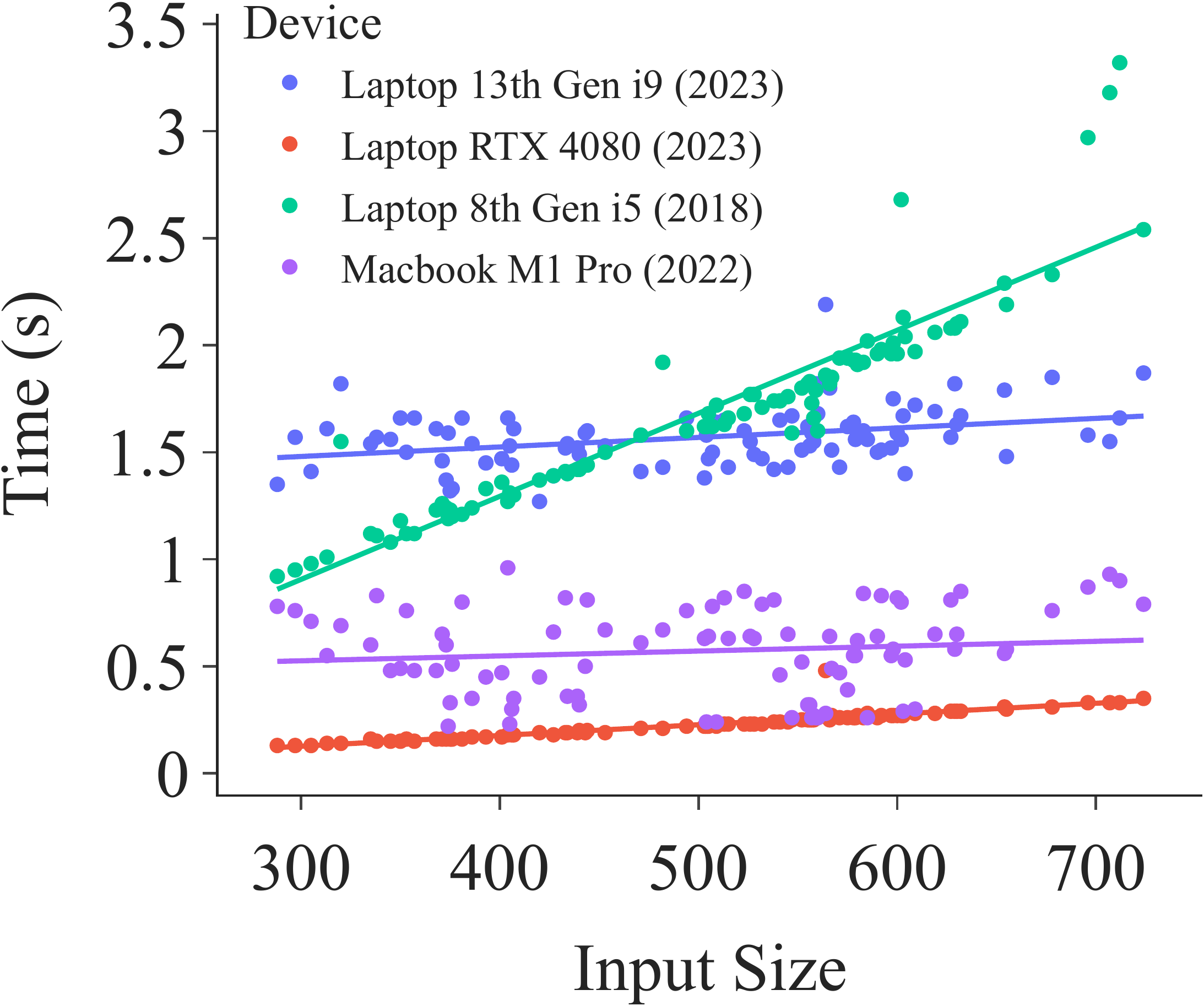}
        \caption{T5's latency when running on the three machines.}
        \label{fig:latency_T5}
    \end{subfigure}
    \hfill
    \begin{subfigure}[t]{0.45\columnwidth}
        \centering
        \includegraphics[width=\textwidth]{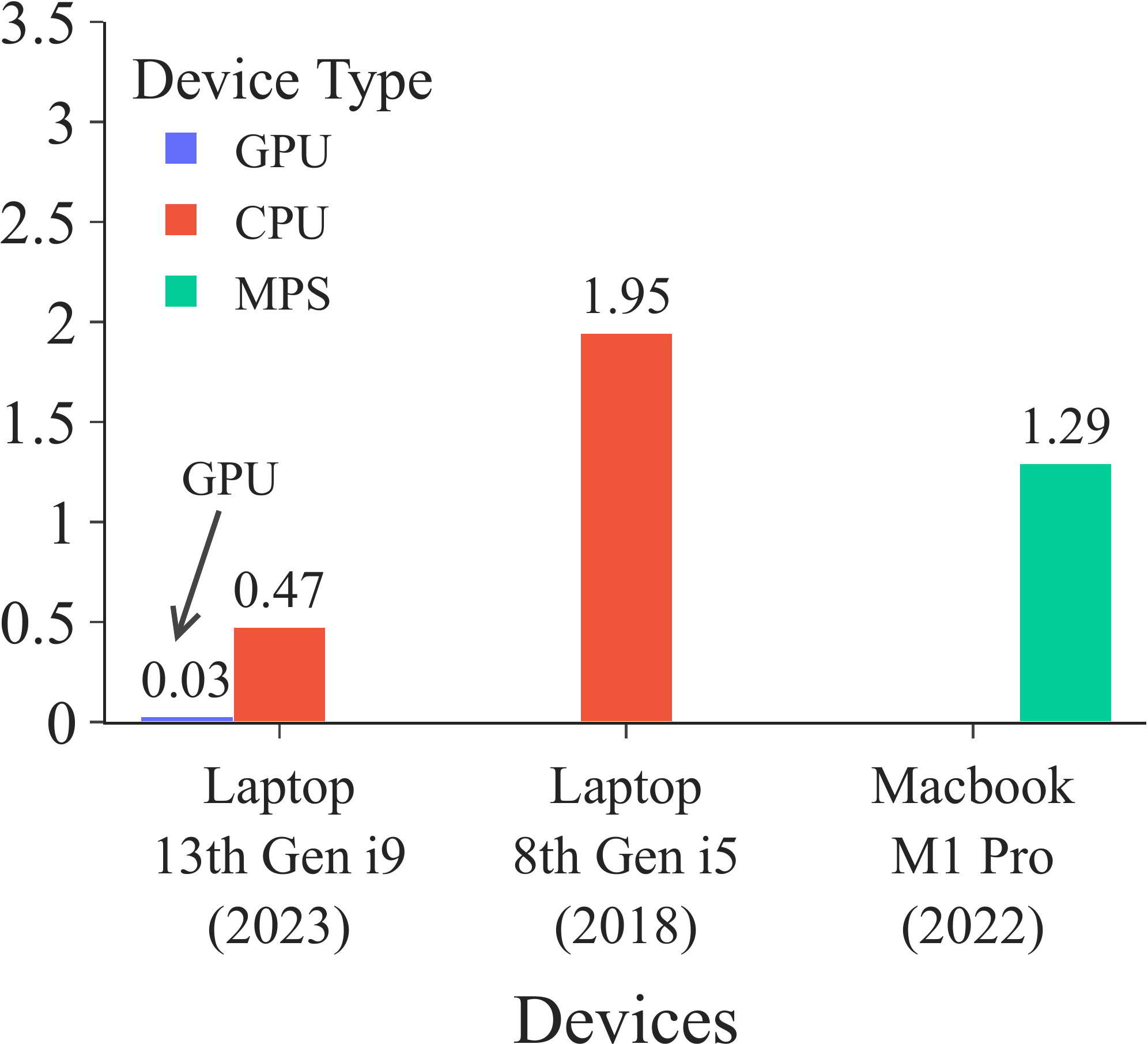}
        \caption{YOLO's latency when running on the three machines}
        \label{fig:latency_yolo}
    \end{subfigure}
    \caption{\texttt{T5} and \texttt{YOLOv10} model inference time running on the 3 machines}
\end{figure}

\newpage

\subsection{Mapping Multi-Token Labels to Single-Token Labels}
\label{app:mapping_multi_single}
\FloatBarrier

\begin{table}[H]
\caption{Mapping between multi-token and single-token representations of dark patterns and user interface concepts}
\centering
\resizebox{\columnwidth}{!}{
\begin{tabular}{llcll}
\hline
\textbf{Label} & \textbf{Multi Tokens} & $\rightarrow$ & \textbf{Single Token} & \textbf{Label} \\
\hline
\rowcolor{aliceblue}
interface-interference & [3459, 18, 3870, 11788, 1] & $\rightarrow$ & [18805, 1] & distraction \\ \midrule
forced-action & [5241, 18, 4787, 1] & $\rightarrow$ & [10472, 1] & obligation \\ \midrule
\rowcolor{aliceblue} 
obstruction & [26359, 1] & $\rightarrow$ & [12515, 1] & barrier \\ \midrule
sneaking & [14801, 53, 1] & $\rightarrow$ & [14801, 1] & sneak \\ \midrule
\rowcolor{aliceblue}
non-deceptive & [529, 18, 221, 6873, 757, 1] & $\rightarrow$ & [26213, 1] & irrelevant \\ \midrule
confirmshaming & [3606, 7, 1483, 53, 1] & $\rightarrow$ & [12447, 1] & shame \\ \midrule
\rowcolor{aliceblue}
fake-scarcity-fake-urgency & [9901, 18, 7, 1720, 6726..., 1] & $\rightarrow$ & [9554, 1] & manufactured \\ \midrule
nudge & [3, 29, 13164, 1] & $\rightarrow$ & [3292, 1] & push \\ \midrule
\rowcolor{aliceblue}
hard-to-cancel & [614, 18, 235, 18, 1608, 7125, 1] & $\rightarrow$ & [19885, 1] & sticky \\ \midrule
pre-selection & [554, 18, 7, 15, 12252, 1] & $\rightarrow$ & [356, 1] & set \\ \midrule
\rowcolor{aliceblue}
visual-interference & [3176, 18, 3870, 11788, 1] & $\rightarrow$ & [21634, 1] & obscure \\ \midrule
jargon & [3, 5670, 5307, 1] & $\rightarrow$ & [11100, 1] & mystery \\ \midrule
\rowcolor{aliceblue}
hidden-subscription & [5697, 18, 7304, 11830, 1] & $\rightarrow$ & [23808, 1] & conceal \\ \midrule
hidden-costs & [5697, 18, 11290, 7, 1] & $\rightarrow$ & [594, 1] & price \\ \midrule
\rowcolor{aliceblue}
disguised-ads & [31993, 26, 18, 9, 26, 7, 1] & $\rightarrow$ & [6543, 1] & ads \\ \midrule
trick-wording & [7873, 18, 6051, 53, 1] & $\rightarrow$ & [21050, 1] & uncertain \\ \midrule
\rowcolor{aliceblue}
not-applicable & [59, 18, 27515, 75, 179, 1] & $\rightarrow$ & [26213, 1] & irrelevant \\
\bottomrule
\end{tabular}
}
\label{app:token_mapping}
\end{table}


\subsection{Taxonomy Mapping from \textit{DPGuard} to \name}
\label{app:dpguard-mapping}
\FloatBarrier

\begin{table}[ht]
    \centering
    \caption{Mapping taxonomy of DPGuard. The table shows deceptive patterns from DPGuard and the deceptive pattern Category/Sub-Type of \name.}
    \resizebox{\columnwidth}{!}{
    \begin{tabular}{lll}
    \textbf{Deceptive Pattern} & \textbf{Category} & \textbf{Sub-Type} \\
    \toprule
    no dp & \revision{Non Deceptive} & \revision{Not Applicable} \\
    nagging & -- & -- \\
    roach motel & -- & -- \\
    price comparison prevention & \revision{Sneaking} & \revision{Hidden Subscription} \\
    intermediate currency & \revision{Forced Action} & \revision{Forced Action} \\
    forced continuity & \revision{Sneaking} & \revision{Hidden Subscription} \\
    hidden costs & \revision{Sneaking} & \revision{Hidden Costs} \\
    sneak into basket & \revision{Sneaking} & \revision{Hidden Subscription} \\
    hidden information & \revision{Obstruction} & \revision{Visual Interference} \\
    preselection & \revision{Obstruction} & \revision{Pre-Selection} \\
    toying with emotion & \revision{Interface Interference} & \revision{Confirmshaming} \\
    false hierarchy & \revision{Interface Interference} & \revision{Nudge} \\
    disguised ads & \revision{Sneaking} & \revision{Disguised Ads} \\
    tricked questions & \revision{Sneaking} & \revision{Trick Wording} \\
    small close button & -- & -- \\
    social pyramid & \revision{Forced Action}& \revision{Forced Action}\\
    privacy zuckering & \revision{Forced Action}& \revision{Forced Action}\\
    gamification & \revision{Forced Action}& \revision{Forced Action}\\
    countdown on ads & \revision{Forced Action} & \revision{Forced Action} \\
    watch ads to unlock features or rewards & \revision{Forced Action} & \revision{Forced Action} \\
    pay to avoid ads & \revision{Forced Action}& \revision{Forced Action}\\
    forced enrollment & \revision{Forced Action} & \revision{Forced Action} \\
    \bottomrule
    \end{tabular}
    }
    \label{tab:taxonomy-dpguard}
\end{table}

\FloatBarrier

\subsection{User Study Mock Website Samples}
\label{app:mock-sites}

\begin{figure}[H]
    \centering
    \includegraphics[width=\columnwidth]{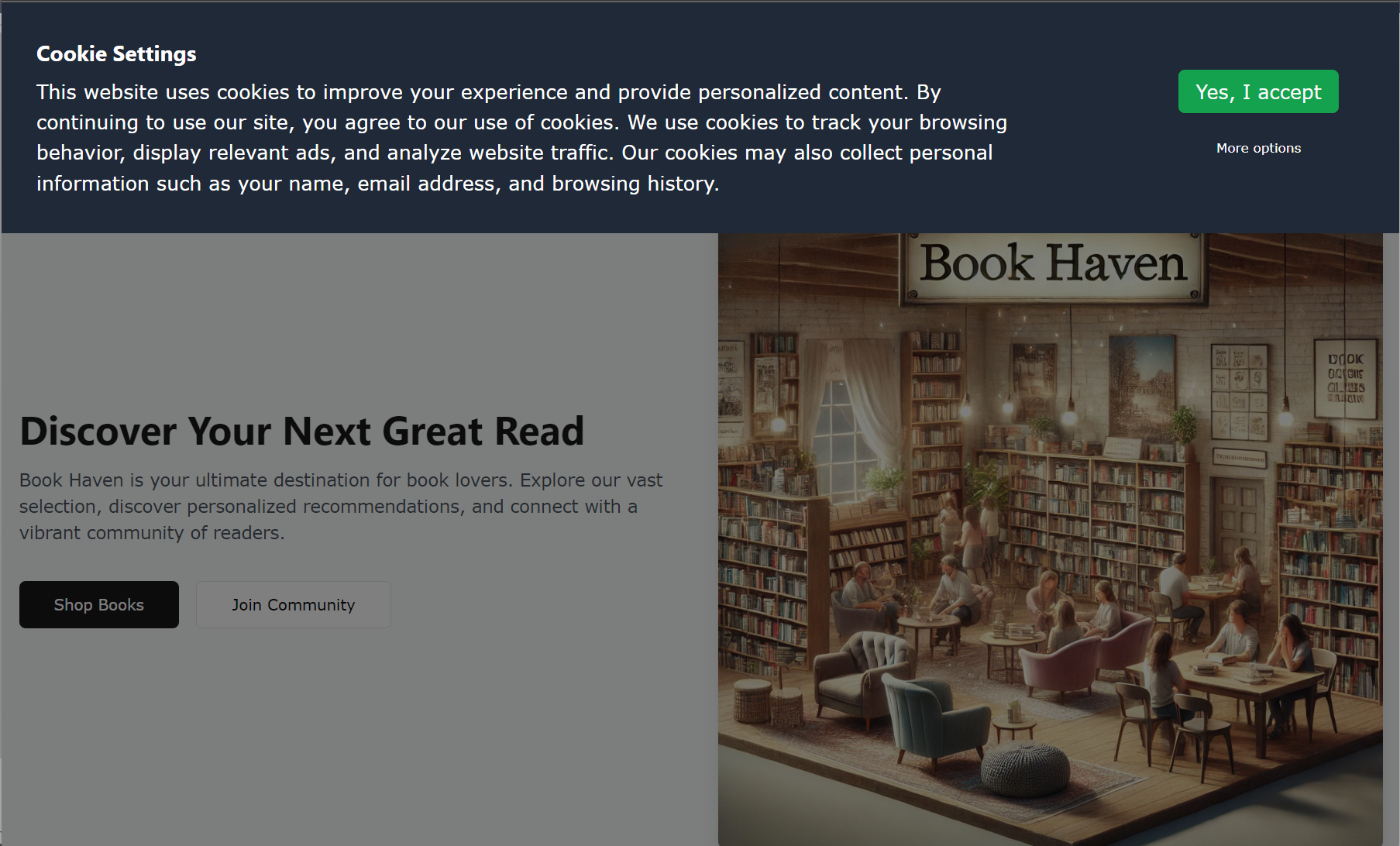}
    \caption{A website showing a book store's homepage with a cookie notice, nudging users to accept cookies.}
    \label{fig:user_study_f1}
\end{figure}
\begin{figure}[H]
    \centering
    \includegraphics[width=\columnwidth]{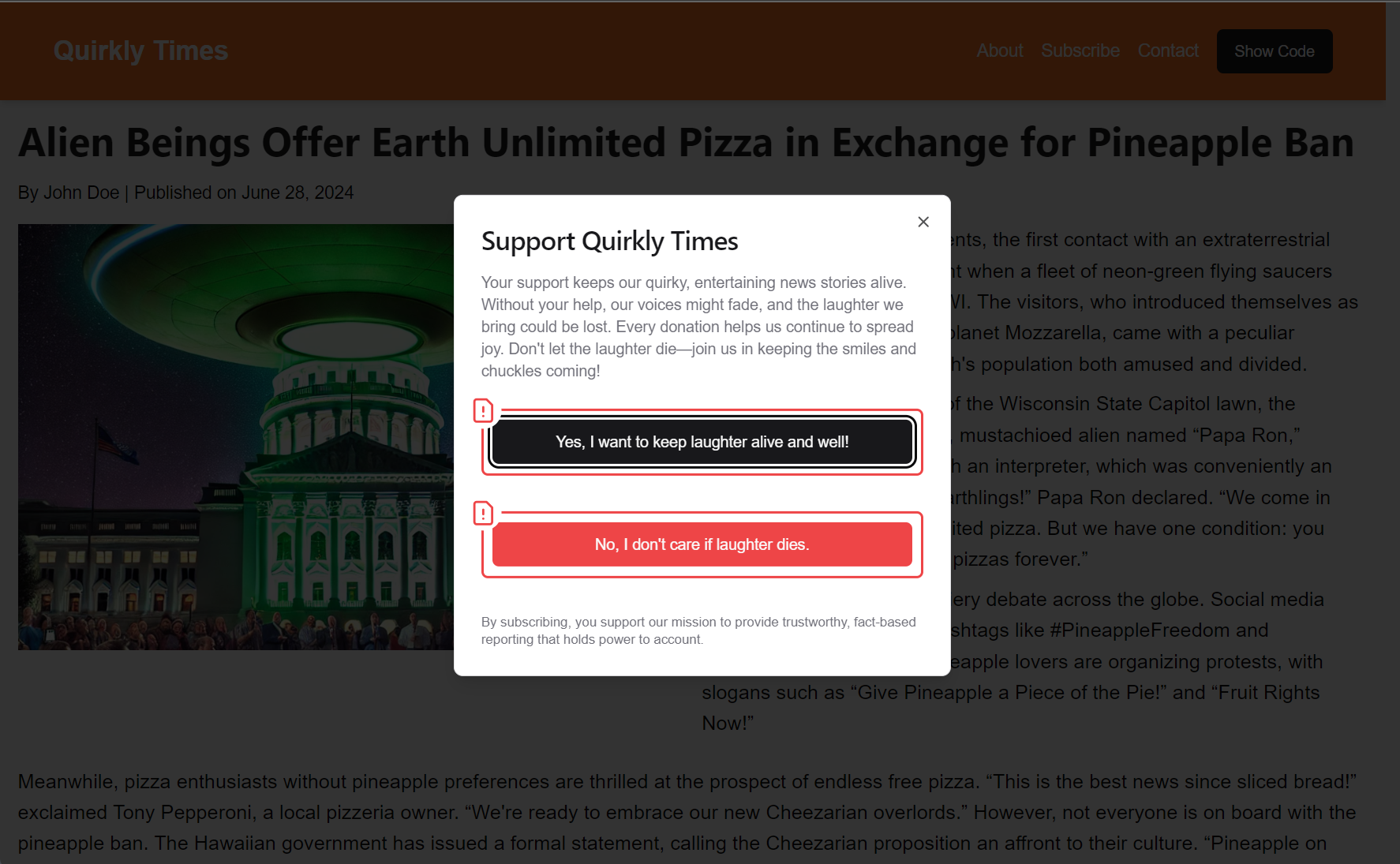}
    \caption{A news organization's website asking users to donate. The deceptive patterns are highlighted in this example with a red box around them.}
    \label{fig:user_study_f2}
\end{figure}

\subsection{User Study Findings}
\label{app:user-study}

\end{document}